\documentclass[useAMS,usenatbib]{mnras}

\pdfoutput=1

\usepackage{amsmath}
\usepackage{bm}
\usepackage{graphicx}
\usepackage{hyperref}
\usepackage[all]{hypcap}
\usepackage{longtable}
\usepackage{multirow}
\usepackage{natbib}
\usepackage{flushend}
\usepackage{txfonts}
\usepackage[capitalize,nameinlink]{cleveref}

\hypersetup{colorlinks,
  linkcolor=red,
  filecolor=red,
  urlcolor=magenta,
  citecolor=blue}

\creflabelformat{equation}{#2#1#3}
\crefrangelabelformat{equation}{#3#1#4 to #5#2#6}

\newcommand\avg[1]{\bar{#1}}
\def\kms{\mathrm{km\,s^{-1}}}
\def\Msun{{\rm M}_\odot}
\def\Nbody{N-body}
\def\percent{ per cent}
\def\sdelta{\mathcal{S}_\Delta}

\def\vbcg{v_{\rm BCG}}
\def\vbcgabs{\lvert\vbcg\rvert}
\def\xcsao{\textsc{xcsao}}

\def\massbias{1.10\pm0.13}

\def\aap{A\&A}
\def\aj{AJ}
\def\ao{AppOpt}
\def\apj{ApJ}
\def\apjl{ApJL}
\def\apjs{ApJS}
\def\apss{Ap\&SS}
\def\araa{ARA\&A}
\def\jcap{JCAP}
\def\mnras{MNRAS}
\def\pasp{PASP}

\def\procspie{SPIE}

\title[Dynamical masses of ACT SZ-selected clusters]
      {The Atacama Cosmology Telescope:
            Dynamical masses for 44 SZ-selected galaxy clusters over 755 square degrees}

\def\leiden{1}
\def\peyton{2}
\def\pennstate{3}
\def\ncsa{4}
\def\illinois{5}
\def\puc{6}
\def\cita{7}
\def\hopkins{8}
\def\upenn{9}
\def\kwazulu{10}
\def\ubc{11}
\def\roma{12}
\def\florida{13}
\def\rutgers{14}
\def\pittsburgh{15}
\def\cornell{16}
\def\jadwin{17}
\def\cmu{18}
\def\goddard{19}
\def\observer{20}

\author[C.\ Sif\'on et al.]
{Crist\'obal~Sif\'on$^{\leiden,\observer}$\thanks{E-mail:sifon@strw.leidenuniv.nl},
Nick~Battaglia$^{\peyton}$,
Matthew~Hasselfield$^{\pennstate,\peyton}$,
Felipe~Menanteau$^{\ncsa,\illinois,\observer}$,
\newauthor
L.~Felipe~Barrientos$^{\puc}$,
J.~Richard~Bond$^{\cita}$,
Devin~Crichton$^{\hopkins}$,
Mark~J.~Devlin$^{\upenn}$,
\newauthor
Rolando~D\"unner$^{\puc}$,
Matt~Hilton$^{\kwazulu}$,
Adam~D.~Hincks$^{\ubc,\roma}$,
Ren\'ee~Hlozek$^{\peyton}$,
\newauthor
Kevin~M.~Huffenberger$^{\florida}$,
John~P.~Hughes$^{\rutgers,\observer}$,
Leopoldo~Infante$^{\puc}$,
Arthur~Kosowsky$^{\pittsburgh}$,
\newauthor
Danica~Marsden$^{\upenn}$,
Tobias~A.~Marriage$^{\hopkins}$,
Kavilan~Moodley$^{\kwazulu}$,
Michael~D.~Niemack$^{\cornell}$,
\newauthor
Lyman~A.~Page$^{\jadwin}$,
David~N.~Spergel$^{\peyton}$,
Suzanne~T.~Staggs$^{\jadwin}$,
Hy~Trac$^{\cmu}$,
\newauthor
Edward~J.~Wollack$^{\goddard}$
\\
$^{\leiden}$Leiden Observatory, Leiden University, PO Box 9513, NL-2300 RA Leiden, Netherlands\\
$^{\peyton}$Department of Astrophysical Sciences, Peyton Hall, Princeton University, Princeton, NJ 
08544, USA\\
$^{\pennstate}$Department of Astronomy and Astrophysics, Davey Lab, The Pennsylvania State 
University, University Park, PA, 16802, USA\\
$^{\ncsa}$National Center for Supercomputing Applications, University of Illinois at 
Urbana-Champaign, 1205 W.\ Clark St., Urbana, IL 61801, USA\\
$^{\illinois}$University of Illinois at Urbana-Champaign, Department of Astronomy, 1002 W.\ Green 
Street, Urbana, IL 61801, USA\\
$^{\puc}$Instituto de Astrof\'isica, Facultad de F\'isica, Pontificia Universidad Cat\'olica, 
Casilla 306, Santiago 22, Chile\\
$^{\cita}$Canadian Institute for Theoretical Astrophysics, 60 St.\ George, Toronto, ON M5S 3H8, 
Canada\\
$^{\hopkins}$Department of Physics and Astronomy, The Johns Hopkins University, 3400 N.\ Charles 
St., Baltimore, MD 21218-2686, USA\\
$^{\upenn}$Department of Physics and Astronomy, University of Pennsylvania, 209 South 33rd Street, 
Philadelphia, PA, 19104 USA\\
$^{\kwazulu}$Astrophysics and Cosmology Research Unit, School of Mathematics, Statistics and 
Computer Science, University of KwaZulu-Natal, Durban 4041, South Africa\\
$^{\ubc}$Department of Physics and Astronomy, University of British Columbia, 6224 Agricultural 
Road, Vancouver, BC V6T 1Z1, Canada\\
$^{\roma}$Pontificia Universit\`a Gregoriana, Piazza della Pilotta 4, 00187 Roma, Italy\\
$^{\florida}$Department of Physics, Florida State University, PO Box 3064350, Tallahassee, FL 
32306-4350, USA\\
$^{\rutgers}$Department of Physics and Astronomy, Rutgers, The State University of New Jersey, 136 
Frelinghuysen Road, Piscataway, NJ 08854-8019, USA\\
$^{\cornell}$Department of Physics, Cornell University, Ithaca, NY, 14853 USA\\
$^{\pittsburgh}$Department of Physics and Astronomy, University of Pittsburgh, Pittsburgh, PA 
15260, USA\\
$^{\jadwin}$Joseph Henry Laboratories of Physics, Jadwin Hall, Princeton University, Princeton, NJ, 
08544, USA\\
$^{\cmu}$McWilliams Center for Cosmology, Department of Physics, Carnegie Mellon University, 
Pittsburgh, PA 15213, USA\\
$^{\goddard}$NASA/Goddard Space Flight Center, Greenbelt, MD 20771, USA
}

\begin{document}

\maketitle

\begin{abstract}
We present galaxy velocity dispersions and dynamical mass estimates for 44 galaxy clusters selected via the Sunyaev-Zel'dovich (SZ) effect by the Atacama Cosmology Telescope. Dynamical masses for 18 clusters are reported here for the first time. Using \Nbody\ simulations, we model the different observing strategies used to measure the velocity dispersions and account for systematic effects resulting from these strategies. We find that the galaxy velocity distributions may be treated as isotropic, and that an aperture correction of up to 7 per cent in the velocity dispersion is required if the spectroscopic galaxy sample is sufficiently concentrated towards the cluster centre. Accounting for the radial profile of the velocity dispersion in simulations enables consistent dynamical mass estimates regardless of the observing strategy. Cluster masses $M_{200}$ are in the range $(1-15)\times10^{14}M_\odot$. Comparing with masses estimated from the SZ distortion assuming a gas pressure profile derived from X-ray observations gives a mean SZ-to-dynamical mass ratio of $\massbias$, but there is an additional 0.14 systematic uncertainty due to the unknown velocity bias; the statistical uncertainty is dominated by the scatter in the mass-velocity dispersion scaling relation. This ratio is consistent with previous determinations at these mass scales.
\end{abstract}


\pagebreak
\section{Introduction}\label{s:intro}

\footnotetext[\observer]{Visiting Astronomer, Gemini South Observatory.}

Galaxy clusters are a sensitive probe of cosmology. Populating the high-end of the mass function, their number density depends strongly on the matter density in the Universe, $\Omega_{\rm m}$, and the amplitude of matter fluctuations, $\sigma_8$ \citep[see, e.g., the review by][]{allen11}. Their potential as cosmological probes, however, depends critically on our knowledge of survey selection effects and baryon physics. Survey selection effects are usually properly accounted for through analytical considerations \citep[e.g.,][]{vikhlinin09_cosmo}, numerical simulations \citep[e.g.,][]{sehgal11,sifon13}, or modeled self-consistently with scaling relations and cosmological parameters \citep[e.g.,][]{pacaud07,mantz10,rozo10,benson13,hasselfield13,bocquet15}. In contrast, incomplete knowledge of baryonic physics poses a serious and still not well understood challenge to the accuracy with which galaxy clusters can constrain cosmological parameters, and is currently the dominant systematic effect \citep[e.g.,][]{benson13,hasselfield13}.

The Sunyaev-Zel'dovich (SZ) effect \citep{zeldovich69,sunyaev80} is a distortion in the cosmic microwave background (CMB) temperature produced by inverse-Compton scattering of CMB photons by free electrons in the hot ($T>10^7\,{\rm K}$) intracluster medium (ICM) of a galaxy cluster. The SZ effect has a distinct frequency dependence such that, in the direction of a massive cluster, the temperature of the sky increases at frequencies larger than 218 GHz while below this frequency the temperature decreases. The amplitude of this distortion is described by the line-of-sight--integrated Compton parameter, $y\propto n_eT_e$, or its solid-angle integral, $Y=\int y\,{\rm d}\Omega$. Its surface brightness is independent of redshift which, to first order, means that surveying the sky at millimetre wavelengths reveals all clusters above a fixed mass to high redshift, resulting in a relatively simple selection function.

Both numerical simulations \citep{springel01_sz,dasilva04,motl05,nagai06,battaglia12} and analytical studies \citep{reid06,afshordi08,shaw08} predict that the SZ effect should correlate with mass with low (of order 10 per cent) intrinsic scatter, although observations correlating the SZ effect with different mass proxies from X-rays \citep{bonamente08,andersson11,planck11xi,benson13,rozo14_scalings}, optical richness \citep{high10,planck11xii,menanteau13,sehgal13}, weak lensing \citep{hoekstra12,marrone12,planckintiii,gruen14} and galaxy velocity dispersion \citep{sifon13,ruel14,rines16} find a larger intrinsic scatter between mass and $Y$ of about 20 per cent. The effect of cluster physics mentioned above, coupled to systematic effects arising from the use of different instruments \citep{mahdavi13,rozo14_observables}, dominate the uncertainties in these scaling relations. This uncertainty has been most notoriously highlighted by the tension in inferred cosmological parameters between the primary CMB and SZ cluster counts found by the Planck satellite \citep{planck15xxiv}, and can be reduced by larger, more detailed analyses involving independent mass proxies and ICM tracers.

Velocity dispersions have been well studied as a proxy for galaxy cluster mass, dating back to the first such scaling relation reported by \cite{evrard89}, and are independent of the ICM properties that determine the SZ effect.\footnote{Some degree of correlation may still exist, however, because different observables are affected by the same large scale structure \citep{white10}.} Extensive tests on numerical simulations have shown that the 3-dimensional galaxy velocity dispersion is a low-scatter mass proxy but, not surprisingly, projection effects including cluster triaxiality and large-scale structure significantly increase the scatter \citep{white10,saro13}. The scatter at fixed velocity dispersion in observed samples is as large as a factor two \citep{old14,old15}.\footnote{In theory, the caustic technique provides a lower-scatter mass proxy than simple velocity dispersions \citep{gifford13}; however, it has been shown to produce similar scatter in more realistic settings \citep{old14,old15}. In this respect, machine learning algorithms may become a promising alternative \citep{ntampaka15,ntampaka16}.} Importantly, the biases on these measurements (typically $\lesssim25$ per cent for $\gtrsim30$ observed galaxies) are much smaller than the observed scatter \citep{old15}, meaning that velocity dispersions remain a valuable, unbiased mass calibrator for sufficiently large cluster samples. In this paper we make use of spectroscopic data to estimate line-of-sight galaxy velocity dispersions (referred to as $\sigma$ in the remainder of this section) and dynamical masses of galaxy clusters selected through their SZ effect using the Atacama Cosmology Telescope \citep[ACT,][]{marriage11_sz,hasselfield13}.

In \cite{sifon13}, we used the $\sigma-M$ scaling relation of \cite{evrard08} to estimate 
dynamical masses of a subset of these clusters. \cite{evrard08} calibrated this scaling relation using a suite of \Nbody\ simulations, using dark matter particles to estimate velocity dispersions. They showed that the velocity dispersions of dark matter particles in \Nbody\ simulations are robust to variations in cosmology and to different simulation codes. However, galaxies, which are used as observational tracers to measure the velocity dispersion, do not necessarily sample the same velocity distribution as the dark matter particles. Both galaxies and dark matter subhaloes (the analogues of galaxies in \Nbody\ simulations) feel dynamical friction, which distorts their velocity distribution and biases their dispersion with respect to dark matter particles. Additionally, subhaloes are tidally stripped and disrupted such that they can drop below the subhalo identification limit of a particle simulation. The lower-velocity subhaloes are more likely to be disrupted, thus the surviving subhaloes have a larger velocity dispersion which again biases the velocity dispersion of subhaloes \citep[e.g.,][]{faltenbacher06}.

The result of these effects is referred to as velocity bias, denoted $b_{\rm v}\equiv\sigma_{\rm gal}/\sigma_{\rm DM}$ \citep[e.g.,][]{carlberg94,colin00}. Baryonic effects are significant when quantifying the amplitude of $b_{\rm v}$: recent high-resolution hydrodynamical simulations show significant differences in the velocity dispersions of subhaloes versus DM particles (roughly +7 per cent, which translates to a $\sim20$ per cent bias in mass), but comparatively little difference between galaxies and dark matter subhaloes \citep[e.g.,][]{lau10,munari13,wu13}.\footnote{Selecting galaxies by stellar mass instead of total mass reduces the strength of the velocity bias \citep[e.g.,][]{faltenbacher06,lau10}.} Additionally, the amplitude of $b_{\rm v}$ depends on the brightness of the observed galaxies: the velocity dispersion of brighter galaxies is generally biased low, but this can be counteracted by selecting a sample of ($\gtrsim30$) galaxies with a representative brightness distribution \citep{old13,wu13}. In apparent contradiction with this, \cite{guo15_vbias,guo15_clustering} used measurements of the clustering of luminous red galaxies (LRGs) to infer a \emph{negative} velocity bias for satellite galaxies. Moreover, they found that \emph{less} luminous LRGs have a stronger velocity bias of about 90 per cent, while more luminous LRGs have velocities consistent with those of DM particles. This result can be reconciled with those of the above simulations by noting that any given cluster\footnote{Note that both the simulations and the observations of \cite{guo15_vbias,guo15_clustering} refer to clusters with masses well below $10^{15}\Msun$.} typically has less than ten LRGs---both \cite{old13} and \cite{wu13} find that taking the $N$ brightest galaxies gives rise to a velocity bias of roughly 0.9, if $N\lesssim10$.

Since observationally one uses galaxies to calculate $\sigma$, biases may be introduced if one uses a $\sigma-M$ scaling relation calibrated from simulations using dark matter particles, such as that of \cite{evrard08}, but does not account for the aforementioned complexities. Therefore, in this paper we use the scaling relation of \cite{munari13}, calibrated on simulated \textit{galaxies} instead of dark matter particles, to relate velocity dispersions to cluster masses.

We present our SZ-selected cluster sample and describe the observations, data reduction and 
archival compilation in \Cref{s:data}. In \Cref{s:masses} we describe our velocity dispersion and 
dynamical mass estimates, including an assessment of our different observing strategies using mock 
observations on numerical simulations (\Cref{s:sims}), a comparison to SZ-derived masses 
(\Cref{s:msz}) and an investigation of cluster substructure (\Cref{s:substruct}). We highlight 
interesting individual clusters in \Cref{s:individual} and summarize the main results in 
\Cref{s:conclusions}.

We assume a flat $\Lambda$CDM cosmology\footnote{Assuming Planck-level uncertainties in 
$\Omega_{\rm m}$ and $H_0$ \citep{planck15xiii} introduces a $<5$ per cent difference in the reported 
masses, accounting for their influence on both member selection (through changes in projected 
physical distances) and the adopted scaling relation.} with $\Omega_{\rm m}=0.3$ and 
$H_0=70\,\mathrm{km\,s^{-1}Mpc^{-1}}$. Throughout this work we quote measurements (e.g., masses, 
$M_\Delta$) at a radius $r_\Delta$, within which the average density is $\Delta$ times the 
critical density of the Universe at the corresponding redshift, where $\Delta=\{200,500\}$.

\section{Data and observations}\label{s:data}

In this section we detail the cluster sample, our follow-up observations and data processing, and 
archival data with which we supplement our observations. In summary, we study 44 SZ-selected 
clusters, of which 28 are in the celestial equator and are the focus of this paper, and 16 
clusters are part of the southern survey and were studied in \cite{sifon13}. We summarize our 
observing runs and sources of archival data in \Cref{t:observations}.

\subsection{The Atacama Cosmology Telescope}

The Atacama Cosmology Telescope (ACT) is a 6-meter off-axis Gregorian telescope located at an 
altitude of 5200 m in the Atacama desert in Chile, designed to observe the CMB at arcminute 
resolution. Between 2007 and 2010, ACT was equipped with three 1024-element arrays of transition 
edge sensors operating at 148, 218, and 277 GHz \citep{fowler07,swetz11}, although only the 148 
GHz band has been used for cluster detection. In this period, ACT observed two regions of the sky, 
one covering 455 sq.\ deg.\ to a typical depth of 60 $\mu$K centred around declination $-53^\circ$ 
\citep[the ``southern'' survey,][]{marriage11_sz,marriage11_sources}, and one covering 504 sq.\ deg.\ 
around the celestial equator, with a typical depth of 44 $\mu$K \citep[the ``equatorial'' 
survey,][]{hasselfield13}. For details on the observational strategy of ACT and map making 
procedure see \cite{dunner13}.

In the remainder of this section we describe ACT detections and follow-up observations of clusters 
in the equatorial survey. Details about the detection and optical confirmation of clusters in the 
southern survey can be found in \cite{marriage11_sz} and \cite{menanteau10_act}, respectively. The 
spectroscopic observations are described in \cite{sifon13} and the latest SZ measurements are 
given in \cite{hasselfield13}.

\subsection{ACT SZ-selected clusters in the equator}\label{s:szobs}

Galaxy clusters were detected in the 148 GHz band by matched-filtering the maps with the pressure 
profile suggested by \cite{arnaud10}, fit to X-ray selected local ($z<0.2$) clusters, with varying 
cluster sizes, $\theta_{500}$, from $1.\!\arcmin18$ to $27\arcmin$. A signal-to-noise (S/N) ratio 
map was extracted from each of these matched-filtered maps and all pixels with ${\rm S/N}>4$ were 
considered as cluster candidates. Cluster properties were extracted only from the map with 
$\theta_{500}=5.\!\arcmin9$. The properties depend weakly on the exact shape of the profile as 
discussed in \cite{hasselfield13}.

Because of the complete overlap of ACT equatorial observations with Sloan Digital Sky Survey Data Release 8 \citep[SDSS DR8,][]{sdss8} imaging, {\em all} cluster candidates were assessed with optical data \citep{menanteau13}. With DR8, clusters can be confidently detected up to $z\approx0.5$. Moreover, 270 sq.\ deg.\ of the ACT survey overlap with the deep, co-added SDSS Stripe82 region \citep[S82,][]{annis14}, which allows the detection of the cluster red sequence up to $z\approx0.8$. Confirmed clusters are all those ${\rm S/N}>4$ candidates for which there are at least 15 galaxies within $1\,h^{-1}\,\mathrm{Mpc}$ of the brightest cluster galaxy and with a photometric redshift within $0.045(1+z)$ of the cluster redshift. We additionally targeted candidate high-redshift, high S/N candidates with near infrared $K_s$-band imaging with the ARC 3.5m telescope at the Apache Point Observatory, which allowed us to confirm five additional clusters at $z\gtrsim1$.\footnote{One of these clusters, ACT-CL~J0012.0$-$0046, associated with an overdensity of red galaxies at $z=1.36$ by \cite{menanteau13}, is detected at much lower significance in new, more sensitive SZ observations performed with ACTPol (M.\ Hilton et al., in prep).}

A total of 68 clusters were confirmed, of which 19 (all at $z>0.65$) were new detections. This sample has been divided into three subsamples: a complete sample of clusters within S82 at $z<1$ and with ${\rm S/N}>5$ (the ``cosmological'' sample, containing 15 clusters), a uniform sample of 34 clusters within S82, and an incomplete sample of 19 clusters up to $z\approx0.7$ in the shallower DR8 region. Confirmed clusters in S82 have redshifts up to $z\approx1.3$ (with the aid of near infrared data for the higher redshifts). See \cite{menanteau13} for more details on the optical and infrared confirmation of clusters in the equatorial survey.

\begin{table*}
\centering
\caption{Summary of spectroscopic observations and sources of archival data. The last column lists 
the number of clusters observed in each program. Previously published data have the corresponding 
references. All SDSS clusters have been observed by us in one of the listed programs as well.}
\label{t:observations}
\begin{tabular}{l c l l l l r}
\hline\hline
 Instrument / & \multirow{2}{*}{Semester} & \multirow{2}{*}{Program} & 
\multirow{2}{*}{PI} & \multirow{2}{*}{Data reference} & \multirow{2}{*}{Sample} & 
\multirow{2}{*}{$\mathrm{N_{cl}}$} \\
 Archival source &  &  &  &  & \\
 \hline
 VLT/FORS2 & 2009B & 084.A-0577 & Infante & \citet{sifon13} & South & 3 \\
           & 2010B & 086.A-0425 & Infante & \citet{sifon13} & South & 2 \\
 Gemini/GMOS & 2009B & GS-2009B-Q-2 & Barrientos & \citet{sifon13} & South & 4 \\
             & 2010B & GS-2010B-C-2 & Barrientos/Menanteau & \citet{sifon13} & South & 10 \\
             & 2011B & GS-2011B-C-1 & Barrientos/Menanteau & this work & Equator & 12 \\
             & 2012A & GS-2012A-C-1 & Menanteau & this work & Equator & 8 \\
 SALT/RSS & 2012A & 2012-1-RSA\_UKSC\_RU-001 & Hilton/Hughes & \citet{kirk15} & Equator & 1 \\
          & 2012B & 2012-2-RSA\_UKSC\_RU-001 & Hilton/Hughes & \citet{kirk15} & Equator & 2 \\
          & 2013A & 2013-1-RSA\_RU-001 & Hilton/Hughes & \citet{kirk15} & Equator & 1 \\
          & 2013B & 2013-2-RSA\_RU-002 & Hilton/Hughes & \citet{kirk15} & Equator & 3 \\
 SDSS DR12 & -- & -- & -- & \citet{sdss12} & Equator & 20 \\
 HeCS & -- & -- & -- & \citet{rines13} & Equator & 3 \\
 \multirow{2}{*}{NED} & \multirow{2}{*}{--} & \multirow{2}{*}{--} & \multirow{2}{*}{--} 
& \citet{soucail88}, & \multirow{2}{*}{Equator} & \multirow{2}{*}{1} \\
 &  &  &  & \citet{dressler99} &  &  \\
 \hline
\end{tabular}
\end{table*}

\begin{figure}
 \centerline{\includegraphics[width=3.4in]{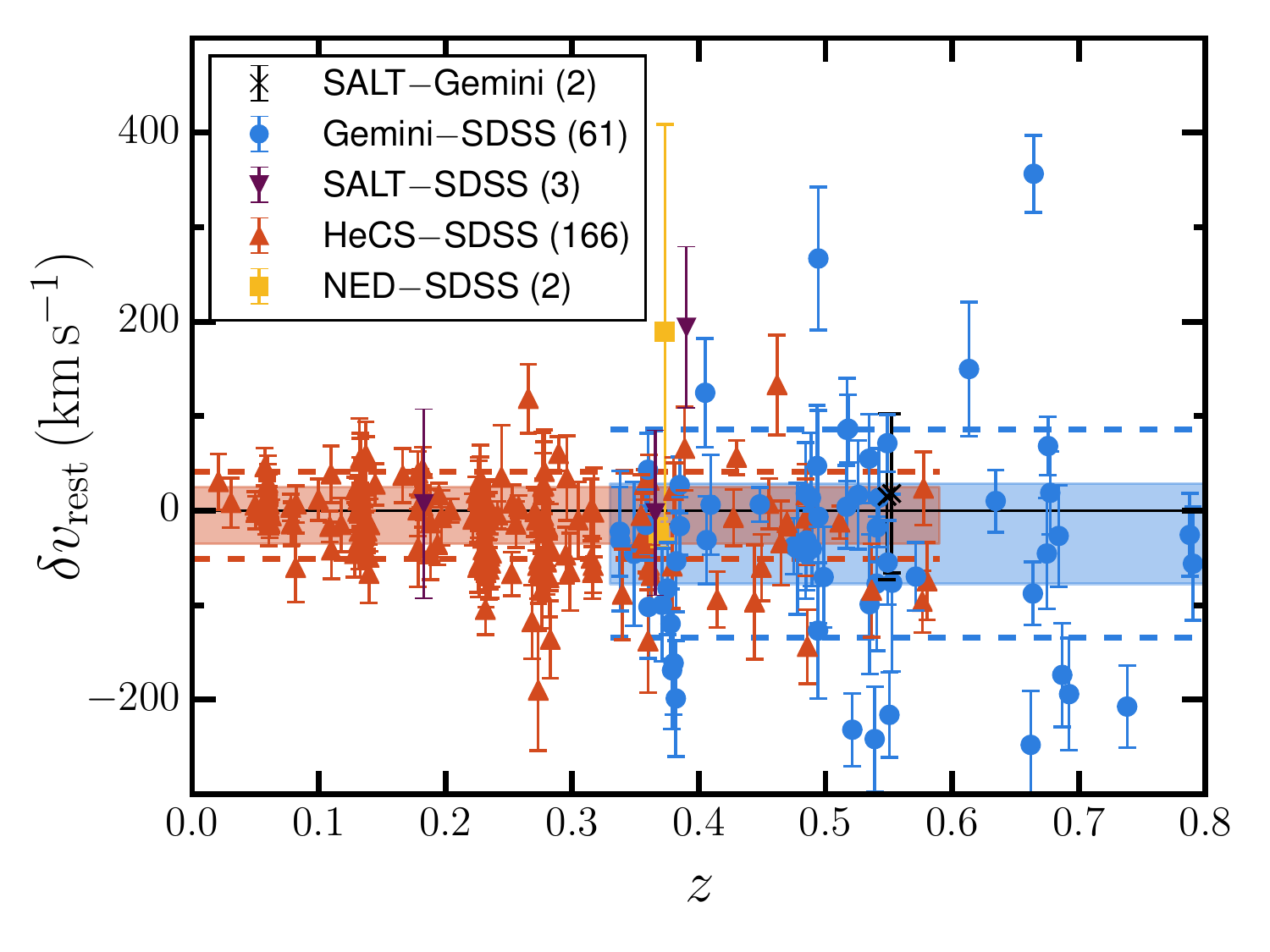}}
\caption{Comparison of redshifts from all spectroscopic datasets to SDSS measurements (for 
overlapping galaxies), shown as $\delta v_{\rm rest}=c\,(z_1-z_2)/(1+z_2)$, where $z_2=z_{\rm 
SDSS}$ (except for the black crosses, where $z_2=z_{\rm Gemini}$). All redshifts are in the 
heliocentric frame. Red, yellow, purple and blue points correspond to redshifts from the HeCS 
survey \citep{rines13}, from NED for Abell~370, and from our SALT/RSS and Gemini/GMOS campaigns, 
respectively, while black crosses compare our redshift measurements between SALT/RSS and 
Gemini/GMOS. Individual uncertainties correspond to the quadrature sum of the uncertainties from 
both measurements. Red and blue shaded regions show uncertainties on the weighted means for 
HeCS$-$SDSS and Gemini$-$SDSS, respectively, and dashed horizontal lines show standard deviations. 
The number of matches per data set pair are given in parentheses in the legend.}
\label{f:redshift_comparison}
\end{figure}

\subsection{Gemini/GMOS spectroscopy}\label{s:spectroscopy}

We observed 20 clusters from the equatorial sample with the Gemini Multi-Object Spectrograph 
\citep[GMOS,][]{hook04} on the Gemini-South telescope, split in semesters 2011B 
(ObsID:GS-2011B-C-1, PI:Barrientos/Menanteau) and 2012A (ObsID:GS-2012A-C-1, PI:Menanteau), 
prioritizing clusters in the cosmological sample at $0.3<z<1.0$. All observations followed our 
setup for the southern sample \citep{sifon13}. We first selected as targets those galaxies with 
photometric redshifts within $\Delta z=0.1$ of the cluster photometric redshift and prioritized 
bright galaxies as allowed by the multi-object spectroscopy (MOS) masks. The only major difference 
in strategy from \cite{sifon13} is that, owing to the SDSS photometry, we targeted galaxies out to 
larger radii than in the southern observations, in which we were bound by the roughly $5\arcmin$ 
fields of view of our targeted optical follow-up with 4m-class telescopes \citep{menanteau10_act}. We 
followed this approach because of the indication, especially from numerical simulations, that the 
velocity dispersion is a decreasing function of radius; therefore an unbiased velocity dispersion 
estimate is predicted only if galaxies are sampled out to approximately the cluster's virial 
radius \citep[e.g.,][]{girardi98,biviano06,mamon10}. We observed 2-3 masks per cluster along the 
(visually identified) major axis of the galaxy distribution. In order to obtain a wide sky 
coverage, these masks were mostly non-overlapping in the sky, even though this meant we had fewer 
targets per unit area. We detail the differences with the southern strategy, and address the 
impact of these differences in our measurements, in \Cref{s:sims}.

We used \texttt{pygmos}\footnote{\url{http://www.strw.leidenuniv.nl/\~sifon/pygmos/}} 
\citep{sifon13}, an automated Python/PyRAF script\footnote{PyRAF is a product of the Space 
Telescope Science Institute, which is operated by AURA for NASA.} that goes from raw data to 
one-dimensional spectra, including bias subtraction, flat field correction, wavelength 
calibration, cosmic ray rejection \citep{vandokkum01} and sky subtraction. Redshifts\footnote{All 
redshifts presented here are in the heliocentric frame.} were measured by cross-correlating the 
spectra with template spectra from 
SDSS\footnote{\url{http://www.sdss.org/DR7/algorithms/spectemplates/index.html}} using \xcsao\ 
within IRAF's \textsc{rvsao} package\footnote{\url{http://tdc-www.harvard.edu/iraf/rvsao/}} 
\citep{kurtz98}.

\subsection{SALT/RSS spectroscopy}\label{s:salt}

We also observed seven clusters in S82 with the Robert Stobie Spectrograph \citep[RSS,][]{burgh03} 
on the Southern African Large Telescope (SALT), using multi-object spectroscopy. Details of these 
observations are given in \cite{kirk15}. Target selection and redshift measurements were carried 
in a similar, but not identical, fashion to the GMOS observations of the equatorial clusters. The 
data were prepared with \texttt{pysalt} \citep{crawford10}, after which they were reduced with 
standard IRAF\footnote{\url{http://iraf.noao.edu/}} functions. Redshift measurements are also 
obtained with \xcsao. ACT-CL~J0045.2$-$0152 is the only cluster that was observed both with Gemini 
and SALT, but there are only two galaxies in our final catalogue observed with both telescopes.

\subsection{Archival data}\label{s:archival}

In order to enlarge the sample of studied clusters and member galaxies, we also compiled archival data for the equatorial sample. Specifically, we searched the SDSS Data Release 12 \citep[DR12,][]{sdss12} database\footnote{\url{http://skyserver.sdss9.org/public/en/home.aspx}}. We retrieved all galaxies with a valid redshift (that is, \texttt{z>0} and \texttt{zWarning=0}) within a cluster-centric distance of $20\arcmin$ (corresponding to several times $r_{200}$ for most clusters) and found a total of 2001 galaxies (most of which are not cluster members; see \Cref{s:masses}) in the direction of 25 of the ACT equatorial clusters observed with Gemini or SALT. Of the galaxies with SDSS spectra, 61 were also observed by us with Gemini, and three with SALT. We compare these repeat observations in \Cref{s:compare_measurements}. There are additionally four clusters in the equatorial sample with dedicated archival observations; we did not observe any of these clusters ourselves. We briefly describe these data below.

The Hectospec Cluster Survey \citep[HeCS,][]{rines13} was designed to measure the masses of galaxy 
clusters at $0.1<z<0.3$ out to the infall regions of clusters (typically around $4r_{200}$), 
targeting more than four hundred objects per cluster within a radius of $30\arcmin$ (corresponding 
to 6 Mpc at $z=0.2$). The three clusters below $z=0.3$ in the cosmological sample of 
\cite{hasselfield13} were targeted by \cite{rines13}, namely ACT-CL J0152.7+0100 (Abell 267), 
ACT-CL J2129.6+0005 (RX J2129.6+0005) and ACT-CL J2337.6+0016 (Abell 2631). We include these three 
clusters in our analysis. \cite{rines13} also measured redshifts using \xcsao; we use only 
galaxies with redshift quality flags \texttt{'Q'} or \texttt{'?'}, which correspond to secure 
redshifts for high- and medium-quality spectra, respectively \citep{rines13}.

Additionally, the cluster ACT-CL J0239.8$-$0134 ($z=0.375$) is the well-studied, HST Frontier 
Fields\footnote{\url{http://www.stsci.edu/hst/campaigns/frontier-fields/}} cluster Abell 370. 
Despite extensive lensing studies 
\citep[e.g.][]{medezinski10,richard10,hoekstra12,vonderlinden14_lensing}, there is no modern spectroscopic 
data on this cluster. A search in the NASA/IPAC Extragalactic 
Database\footnote{\url{http://ned.ipac.caltech.edu}} (NED) gives roughly 100 galaxies with 
redshifts in the range $0.30 \leq z \leq 0.45$, which safely includes all potential cluster 
members (we then run our membership algorithm on these galaxies, see \Cref{s:vdisp}). These 
galaxies go out to $6\arcmin$ in radius. For homogeneity, we limit ourselves to redshifts measured 
either by \cite{soucail88} or \cite{dressler99} since these two sources make up the majority 
($\approx90$ per cent) of galaxies returned by NED. We assign to each galaxy an uncertainty at the 
level 
of the last non-zero digit.

\subsection{Comparison between redshift measurements}
\label{s:compare_measurements}

There are many overlapping galaxies between SDSS and other data, as well as two overlapping 
galaxies between our SALT/RSS and Gemini/GMOS observations of ACT-CL~J0045.2$-$0152. We compare 
the spectroscopic redshifts between the different measurements in \Cref{f:redshift_comparison}. 
There is good agreement between the different datasets. In particular, the 
inverse-variance-weighted average differences in rest-frame velocity (defined as $\delta v_{\rm 
rest}=c\,(z_1-z_2)/(1+z_2)$) are $\delta v_{\rm rest}=-24.2\pm53.1\,\kms$ (where 
the errorbar is the uncertainty on the mean) between GMOS and SDSS, with a standard deviation 
$\sigma_{\delta v}=110\,\kms$, and $\delta v_{\rm 
rest}=-5.1\pm30.0\,\kms$ between HeCS and SDSS, with $\sigma_{\delta 
v}=46\,\kms$. The standard deviations are 2.07 and 1.39 times the average \xcsao\ 
errors, respectively. We conclude that \xcsao\ underestimates the true cross-correlation velocity 
uncertainty by up to a factor two, consistent with previous determinations 
\citep[e.g.,][]{quintana00,boschin04,barrena09}. HeCS spectra have a higher S/N than GMOS spectra; 
therefore it is possible that the level of underestimation depends on the S/N of the spectrum.

\section{Velocity Dispersions and Dynamical Masses}
\label{s:masses}

\subsection{Velocity dispersion measurements}
\label{s:vdisp}

We use the shifting gapper method developed by \cite{fadda96} as implemented in \cite{sifon13} to select cluster members, as follows. Assuming the BCG to correspond to the cluster centre\footnote{The only exception is ACT-CL~J2302.5+0002, which we discuss in \Cref{s:J2302}.} (the impact of this assumption is assessed in \Cref{s:miscentring}), we bin galaxies by their (projected) cluster-centric distance in bins of at least 250 kpc and 10 galaxies. Therefore member selection in clusters with fewer than 20 redshifts was performed using a single bin (i.e., a standard sigma-clipping). A visual inspection of the phase-space diagrams of clusters with few members suggests that this choice is better than the 15 galaxies used in \cite{sifon13}, where clusters had an average 65 members over a smaller area of the sky.\footnote{The difference in the dynamical masses (which are reported in \Cref{s:mdyn}) between using 10 or 15 galaxies as a minimal bin size in the shifting gapper is six per cent, well within the reported errorbars.} In each bin in projected distance we sort galaxies by the absolute value of their peculiar velocity (taken initially with respect to the median redshift of potential cluster members). In practice, this means we assume that clusters are symmetric in the radial direction. We then select a main body of galaxies having peculiar velocities $\lvert v_i \rvert < \lvert v_{i-1} \rvert +500\,\kms$, where the index $i$ runs over all galaxies in a given radial bin. In other words, the main body is composed, in each radial bin, by the group of galaxies intersecting $v=0$ and bound by velocity differences of less than 500 $\kms$. All galaxies with peculiar velocities less than $1000\,\kms$ away from the main body are considered cluster members. Modifying the velocity gaps does not have a noticeable impact on our results---all clusters have well defined boundaries in velocity space. This process is iterated, updating the cluster redshift and the radial binning, until the number of members converges (usually two to three iterations).

At every step in the member selection process, cluster redshifts and velocity dispersions are calculated as the biweight estimators of location and scale \citep[i.e., the central value and dispersion, respectively, see equations 5 and 9 of][]{beers90}, respectively. We correct the velocity  dispersion for individual redshift uncertainties \citep{danese80}, but this is a $<1$ per cent correction for $\sigma=1000\,\kms$. We estimate 68 per cent uncertainties in cluster redshifts and velocity dispersions by bootstrapping over all galaxies within $3\sigma$ of the measured velocity dispersion, which is always larger than the velocity limit defined by the shifting gapper. Therefore we include galaxies which are rejected by our member selection algorithm, and thus account for uncertainties arising from membership selection in the redshift and velocity dispersion uncertainties. We find that the membership selection process increases the statistical uncertainties in the mass by a median 2 per cent for the full sample (but by $>20$ per cent for nine clusters where a large number of objects are rejected by the member selection algorithm). Such a small value is dominated by the southern clusters where, since we targeted the central regions only, the number of galaxies rejected by our algorithm is small compared to the number of members (only 2 per cent, compared to 24 per cent of galaxies rejected for the equatorial clusters). For comparison, we also implement a Bayesian algorithm to estimate velocity dispersions statistically accounting for an interloper component with constant spatial density \citep{wojtak07,andreon08}. The Bayesian analysis yields velocity dispersions, as well as uncertainties, that are consistent with our analysis.

Cluster redshifts and velocity dispersions are listed in \Cref{t:masses}. We show velocity histograms of clusters in the equatorial and southern samples in \Cref{f:hist_eq,f:hist_south}, respectively.

\begin{table*}
\begin{minipage}{\textwidth}
\centering
\caption{Redshifts, velocity dispersions and dynamical masses of ACT SZ-selected clusters. The 
horizontal line separates equatorial and southern clusters. Clusters in the cosmological samples 
of \citet{hasselfield13} have a ``Cosmo'' suffix in the second column. The third and fourth 
columns give the total number of members, $N_{\rm m}$, and the number of members within $r_{200}$, 
$N_{200}$. We list the maximum radius at which we have spectroscopic members, $r_{\rm max}$, and 
the velocity dispersion of all members, $\sigma(<r_{\rm max})$, as well as the quantities 
calculated specifically within $r_{200}$. Uncertainties in the masses do not include the scatter 
in the $\sigma-M$ scaling relation. Alternative cluster names are given in \citet{menanteau13}.}
\label{t:masses}
\begin{tabular}{l c r r c r@{}l c r@{}l c r@{}l}
\hline\hline
Cluster & Sample & $N_{\rm m}$ & $N_{200}$ & $z_{\rm cl}$ & \multicolumn{2}{c}{$\sigma(<r_{\rm 
max})$} & $r_{\rm 
max}$ & \multicolumn{2}{c}{$\sigma_{200}$} & $r_{200}$ & \multicolumn{2}{c}{$M_{200}$} \\
 &  &  &  &  & \multicolumn{2}{c}{$(\kms)$} & $(r_{200})$ & 
\multicolumn{2}{c}{$(\kms)$} & $(\mathrm{Mpc})$ & 
\multicolumn{2}{c}{$(10^{14}\,\Msun)$} \\[0.5ex]
\hline
ACT-CL~J0014.9$-$0057 & S82-Cosmo & 62 & 45 & $0.5331\pm0.0007$ & 806\, & $\pm\,91$ & 1.75 & 850\, 
& $\pm\,108$ & $1.31\pm0.16$ & 4.5\, & $\pm\,1.6$ \\
ACT-CL~J0022.2$-$0036 & S82-Cosmo & 55 & 44 & $0.8048\pm0.0014$ & 961\, & $\pm\,124$ & 1.71 & 
1025\, & $\pm\,164$ & $1.33\pm0.20$ & 6.6\, & $\pm\,3.0$ \\
ACT-CL~J0045.2$-$0152 & DR8 & 56 & 44 & $0.5483\pm0.0010$ & 930\, & $\pm\,77$ & 1.40 & 967\, & 
$\pm\,88$ & $1.45\pm0.12$ & 6.3\, & $\pm\,1.6$ \\
ACT-CL~J0059.1$-$0049 & S82-Cosmo & 44 & 23 & $0.7870\pm0.0012$ & 884\, & $\pm\,150$ & 1.82 & 
874\, & $\pm\,206$ & $1.19\pm0.28$ & 4.6\, & $\pm\,3.2$ \\
ACT-CL~J0119.9+0055 & S82 & 16 & 14 & $0.7310\pm0.0011$ & 725\, & $\pm\,128$ & 1.06 & 786\, & 
$\pm\,149$ & $1.10\pm0.20$ & 3.4\, & $\pm\,1.9$ \\
ACT-CL~J0127.2+0020 & S82 & 46 & 46 & $0.3801\pm0.0008$ & 994\, & $\pm\,106$ & 0.92 & 991\, & 
$\pm\,108$ & $1.64\pm0.17$ & 7.5\, & $\pm\,2.3$ \\
ACT-CL~J0152.7+0100 & S82-Cosmo & 253 & 144 & $0.2291\pm0.0004$ & 931\, & $\pm\,41$ & 2.57 & 
1065\, & $\pm\,54$ & $1.89\pm0.09$ & 9.7\, & $\pm\,1.4$ \\
ACT-CL~J0206.2$-$0114 & S82-Cosmo & 40 & 23 & $0.6758\pm0.0010$ & 570\, & $\pm\,105$ & 2.02 & 
625\, & $\pm\,164$ & $0.94\pm0.25$ & 2.0\, & $\pm\,1.6$ \\
ACT-CL~J0215.4+0030 & S82-Cosmo & 14 & 11 & $0.8622\pm0.0026$ & 1386\, & $\pm\,262$ & 1.27 & 
1256\, & $\pm\,268$ & $1.57\pm0.33$ & 11.7\, & $\pm\,7.3$ \\
ACT-CL~J0218.2$-$0041 & S82-Cosmo & 61 & 41 & $0.6727\pm0.0008$ & 723\, & $\pm\,76$ & 1.80 & 790\, 
& $\pm\,92$ & $1.12\pm0.12$ & 3.4\, & $\pm\,1.1$ \\
ACT-CL~J0219.9+0129 & DR8 & 10 & 10 & $0.3651\pm0.0014$ & 1001\, & $\pm\,224$ & 0.56 & 963\, & 
$\pm\,215$ & $1.66\pm0.36$ & 7.6\, & $\pm\,5.0$ \\
ACT-CL~J0223.1$-$0056 & S82-Cosmo & 38 & 27 & $0.6632\pm0.0011$ & 829\, & $\pm\,96$ & 1.25 & 911\, 
& $\pm\,165$ & $1.31\pm0.23$ & 5.3\, & $\pm\,2.8$ \\
ACT-CL~J0239.8$-$0134 & DR8 & 75 & 75 & $0.3751\pm0.0009$ & 1216\, & $\pm\,128$ & 0.64 & 1183\, & 
$\pm\,128$ & $1.94\pm0.19$ & 12.2\, & $\pm\,3.7$ \\
ACT-CL~J0241.2$-$0018 & S82 & 36 & 26 & $0.6872\pm0.0013$ & 830\, & $\pm\,132$ & 1.70 & 905\, & 
$\pm\,160$ & $1.28\pm0.22$ & 5.1\, & $\pm\,2.6$ \\
ACT-CL~J0256.5+0006 & S82-Cosmo & 78 & 78 & $0.3625\pm0.0008$ & 1185\, & $\pm\,102$ & 0.59 & 
1144\, & $\pm\,102$ & $1.89\pm0.16$ & 11.2\, & $\pm\,2.8$ \\
ACT-CL~J0320.4+0032 & S82 & 25 & 25 & $0.3847\pm0.0014$ & 1284\, & $\pm\,209$ & 0.56 & 1236\, & 
$\pm\,215$ & $2.03\pm0.34$ & 14.3\, & $\pm\,7.1$ \\
ACT-CL~J0326.8$-$0043 & S82-Cosmo & 62 & 59 & $0.4471\pm0.0006$ & 897\, & $\pm\,96$ & 1.21 & 927\, 
& $\pm\,101$ & $1.49\pm0.15$ & 6.0\, & $\pm\,1.8$ \\
ACT-CL~J0342.7$-$0017 & S82 & 19 & 19 & $0.3072\pm0.0015$ & 941\, & $\pm\,173$ & 0.80 & 930\, & 
$\pm\,173$ & $1.64\pm0.29$ & 6.9\, & $\pm\,3.7$ \\
ACT-CL~J0348.6$-$0028 & S82 & 15 & 15 & $0.3449\pm0.0010$ & 642\, & $\pm\,117$ & 0.51 & 614\, & 
$\pm\,112$ & $1.10\pm0.19$ & 2.2\, & $\pm\,1.1$ \\
ACT-CL~J2050.5$-$0055 & S82-Cosmo & 33 & 14 & $0.6226\pm0.0007$ & 539\, & $\pm\,120$ & 2.32 & 
511\, & $\pm\,97$ & $0.79\pm0.14$ & 1.1\, & $\pm\,0.6$ \\
ACT-CL~J2050.7+0123 & DR8 & 47 & 47 & $0.3339\pm0.0009$ & 1046\, & $\pm\,104$ & 0.92 & 1043\, & 
$\pm\,103$ & $1.76\pm0.16$ & 8.8\, & $\pm\,2.4$ \\
ACT-CL~J2055.4+0105 & S82 & 55 & 52 & $0.4089\pm0.0005$ & 759\, & $\pm\,77$ & 1.11 & 778\, & 
$\pm\,78$ & $1.29\pm0.12$ & 3.8\, & $\pm\,1.1$ \\
ACT-CL~J2058.8+0123 & DR8 & 16 & 16 & $0.3285\pm0.0014$ & 1109\, & $\pm\,196$ & 0.65 & 1080\, & 
$\pm\,191$ & $1.86\pm0.31$ & 10.2\, & $\pm\,5.2$ \\
ACT-CL~J2128.4+0135 & DR8 & 59 & 56 & $0.3856\pm0.0006$ & 895\, & $\pm\,116$ & 1.12 & 906\, & 
$\pm\,119$ & $1.51\pm0.19$ & 5.9\, & $\pm\,2.2$ \\
ACT-CL~J2129.6+0005 & S82-Cosmo & 291 & 68 & $0.2337\pm0.0005$ & 786\, & $\pm\,39$ & 5.05 & 859\, 
& $\pm\,91$ & $1.56\pm0.15$ & 5.5\, & $\pm\,1.6$ \\
ACT-CL~J2154.5$-$0049 & S82-Cosmo & 52 & 42 & $0.4904\pm0.0011$ & 918\, & $\pm\,108$ & 1.42 & 
964\, & $\pm\,121$ & $1.51\pm0.18$ & 6.6\, & $\pm\,2.3$ \\
ACT-CL~J2302.5+0002 & S82 & 47 & 39 & $0.5199\pm0.0007$ & 648\, & $\pm\,67$ & 1.16 & 671\, & 
$\pm\,63$ & $1.06\pm0.11$ & 2.4\, & $\pm\,0.7$ \\
ACT-CL~J2337.6+0016 & S82-Cosmo & 154 & 51 & $0.2769\pm0.0007$ & 853\, & $\pm\,52$ & 4.85 & 879\, 
& $\pm\,96$ & $1.56\pm0.16$ & 5.7\, & $\pm\,1.7$ \\
\hline
ACT-CL~J0102$-$4915 & South-Cosmo & 86 & 81 & $0.8700\pm0.0010$ & 1273\, & $\pm\,114$ & 1.18 & 
1284\, & $\pm\,117$ & $1.55\pm0.13$ & 11.3\, & $\pm\,2.9$ \\
ACT-CL~J0215$-$5212 & South & 54 & 54 & $0.4803\pm0.0009$ & 1027\, & $\pm\,110$ & 0.83 & 1018\, & 
$\pm\,111$ & $1.59\pm0.16$ & 7.6\, & $\pm\,2.3$ \\
ACT-CL~J0232$-$5257 & South & 63 & 63 & $0.5561\pm0.0007$ & 924\, & $\pm\,87$ & 0.66 & 900\, & 
$\pm\,86$ & $1.35\pm0.12$ & 5.2\, & $\pm\,1.4$ \\
ACT-CL~J0235$-$5121 & South & 80 & 80 & $0.2775\pm0.0005$ & 1044\, & $\pm\,93$ & 0.48 & 994\, & 
$\pm\,92$ & $1.74\pm0.15$ & 8.0\, & $\pm\,2.0$ \\
ACT-CL~J0237$-$4939 & South & 65 & 65 & $0.3343\pm0.0007$ & 1290\, & $\pm\,91$ & 0.39 & 1210\, & 
$\pm\,91$ & $2.01\pm0.14$ & 13.1\, & $\pm\,2.7$ \\
ACT-CL~J0304$-$4921 & South & 61 & 61 & $0.3917\pm0.0007$ & 1098\, & $\pm\,98$ & 0.51 & 1050\, & 
$\pm\,96$ & $1.71\pm0.15$ & 8.7\, & $\pm\,2.2$ \\
ACT-CL~J0330$-$5227 & South-Cosmo & 71 & 71 & $0.4417\pm0.0008$ & 1247\, & $\pm\,96$ & 0.46 & 
1182\, & $\pm\,98$ & $1.85\pm0.14$ & 11.6\, & $\pm\,2.7$ \\
ACT-CL~J0346$-$5438 & South & 88 & 88 & $0.5297\pm0.0007$ & 1081\, & $\pm\,76$ & 0.77 & 1066\, & 
$\pm\,75$ & $1.60\pm0.10$ & 8.3\, & $\pm\,1.6$ \\
ACT-CL~J0438$-$5419 & South-Cosmo & 63 & 63 & $0.4212\pm0.0009$ & 1268\, & $\pm\,109$ & 0.56 & 
1221\, & $\pm\,108$ & $1.93\pm0.16$ & 12.9\, & $\pm\,3.2$ \\
ACT-CL~J0509$-$5341 & South-Cosmo & 74 & 71 & $0.4601\pm0.0005$ & 860\, & $\pm\,79$ & 1.13 & 865\, 
& $\pm\,82$ & $1.38\pm0.12$ & 4.9\, & $\pm\,1.3$ \\
ACT-CL~J0521$-$5104 & South & 19 & 19 & $0.6742\pm0.0018$ & 941\, & $\pm\,194$ & 0.97 & 940\, & 
$\pm\,198$ & $1.35\pm0.28$ & 5.9\, & $\pm\,3.6$ \\
ACT-CL~J0528$-$5259 & South & 55 & 44 & $0.7676\pm0.0010$ & 934\, & $\pm\,114$ & 1.45 & 984\, & 
$\pm\,125$ & $1.30\pm0.16$ & 5.9\, & $\pm\,2.1$ \\
ACT-CL~J0546$-$5345 & South-Cosmo & 45 & 40 & $1.0668\pm0.0013$ & 1020\, & $\pm\,138$ & 1.20 & 
1018\, & $\pm\,148$ & $1.13\pm0.15$ & 5.5\, & $\pm\,2.3$ \\
ACT-CL~J0559$-$5249 & South-Cosmo & 25 & 25 & $0.6094\pm0.0016$ & 1085\, & $\pm\,136$ & 0.86 & 
1078\, & $\pm\,137$ & $1.55\pm0.19$ & 8.3\, & $\pm\,3.0$ \\
ACT-CL~J0616$-$5227 & South-Cosmo & 18 & 18 & $0.6837\pm0.0015$ & 1156\, & $\pm\,193$ & 0.75 & 
1139\, & $\pm\,190$ & $1.58\pm0.25$ & 9.5\, & $\pm\,4.5$ \\
ACT-CL~J0707$-$5522 & South & 58 & 58 & $0.2958\pm0.0005$ & 838\, & $\pm\,82$ & 0.66 & 816\, & 
$\pm\,83$ & $1.44\pm0.14$ & 4.6\, & $\pm\,1.3$ \\
\hline
\end{tabular}
\end{minipage}
\end{table*}

\subsection{Calibrating velocity dispersions with the Multidark simulation}
\label{s:sims}

\begin{figure*}
 \centerline{\includegraphics[width=3in]{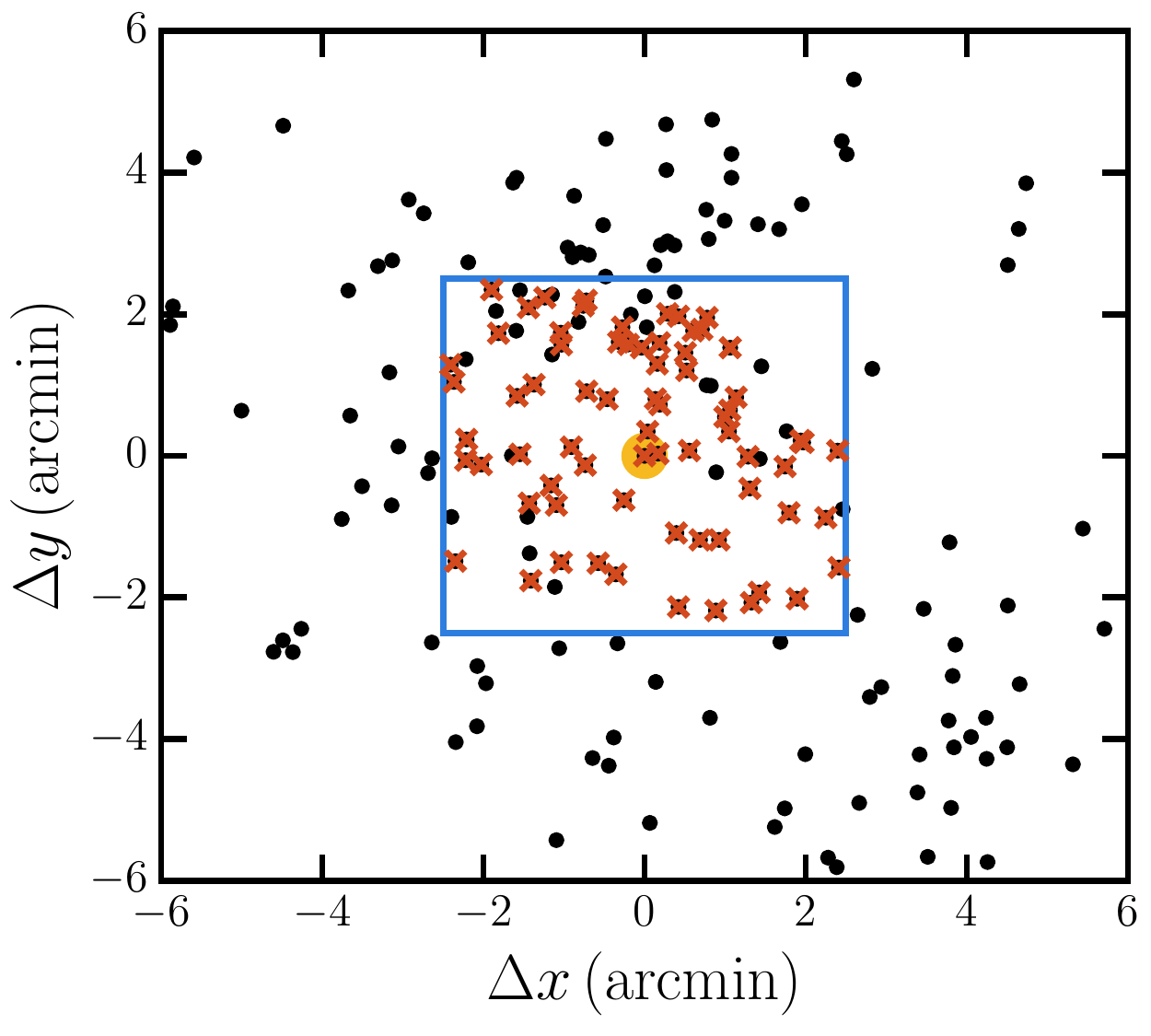}
             \includegraphics[width=3in]{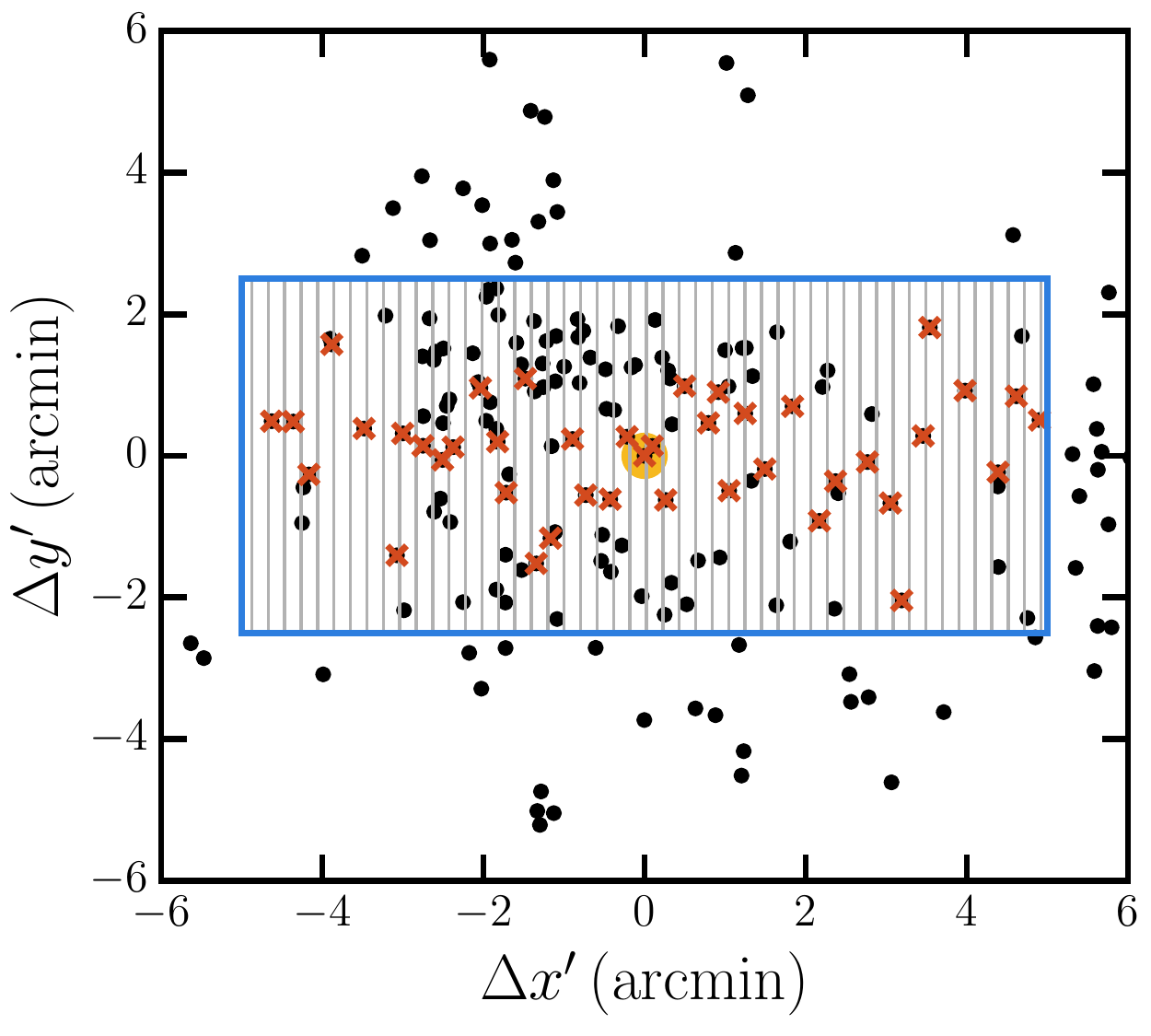}}
\caption{The two simulated observational strategies, for a Multidark halo of mass 
$M_{200}=1.76\times10^{15}\Msun$. The same halo is shown in both panels; in the right panel axes 
are rotated such that the slits are placed along the horizontal axis. Angular distances are
scaled to $z=0.5$; at this redshift, the size $r_{200}$ of this halo corresponds to 
$7.\!\arcmin6.$. Black dots are all halo members, of which red crosses are used to calculate the 
velocity dispersion. The orange circle in the middle marks the central subhalo, which is always 
used to calculate the velocity dispersion, and the blue rectangle outlines the field of view. 
\textit{Left:} the southern strategy, in which we observed up to 70 randomly selected members in 
the central $5\arcmin\times5\arcmin$. \textit{Right:} the equatorial strategy, in which we 
observed an average 25 members inside a $10\arcmin\times5\arcmin$ field of view along the major 
axis of the subhalo distribution. Grey vertical stripes show the mask slit boundaries, and only 
one galaxy is observed per slit.}
\label{f:observations}
\end{figure*}

To estimate a possible bias in the velocity dispersions arising from the different optical observations, especially between our southern and equatorial campaigns, we use mock observations of dark matter haloes in the Multidark simulation \citep{prada12}. Here we want to understand whether there is a relative bias between the two strategies compared to the ``true'' velocity dispersion. By ``true'' we mean the line-of-sight velocity dispersion obtained using all the subhaloes found in the simulation within $r_{200}$, where $r_{200}$ is measured directly in the simulation as the distance from the centre of mass within which the density is 200 times the critical density. We begin by describing the Multidark simulation and then describe our mock observations of subhaloes that follow our real observing strategies.

We used haloes from the Multidark BDMW database \citep{riebe13} constructed from the \Nbody\ 
Multidark MDPL simulation \citep{prada12}. The Multidark simulation is an \Nbody\ simulation 
containing $3840^3$ dissipationless particles in a box of length $1\,h^{-1}{\rm Gpc}$ and run 
using a variation of the \textsc{Gadget2} code \citep{springel05_gadget}. The halo catalog was 
constructed using a spherical over-density halo finder that used the bound density maxima 
algorithm \citep[BDM,][]{klypin97} with an over-density criterion of 200 times the critical 
density of the Universe. The cosmology used in the simulation is a concordance $\Lambda$CDM model 
that is consistent with \cite{planck13xvi}; the parameters are $\Omega_\Lambda=0.69$, $\Omega_{\rm 
m}=0.31$, $\Omega_\mathrm{b}=0.048$, $h=0.68$, and $\sigma_8=0.82$. The small differences in 
cosmological parameters between the simulations and those adopted by us ($\Omega_{\rm m}=0.3$, 
$h=0.7$) have no impact on our results.

We select all haloes at $z=0$ more massive than $10^{14}\,h^{-1}\Msun$ and containing a minimum 
number of 50 subhaloes more massive than $10^{12}\,h^{-1}\Msun$. A total of 572 haloes meet these 
criteria. We created mock observations of the Multidark haloes by implementing distinct algorithms 
for the southern and equatorial strategies to mimic our observational strategies. While the 
southern campaign was confined to areas $\approx\!5\arcmin\times5\arcmin$ around the BCGs, for the 
equatorial sample we tried to observe as far out as possible (see \Cref{s:spectroscopy}). First, 
we scaled projected distances of the subhaloes to $z=0.5$, the median redshift of our sample.
As our sample spans $0.25<z<1.06$, an observing field of fixed angular extent contains different 
fractions of $r_{200}$. However, as we show below, the most important parameter is the radial 
coverage. There is therefore no extra information in scaling distances to different redshifts.

We note that our goal in this section is not to test the membership selection algorithm, and we therefore only include subhaloes in Multidark within $r_{200}$. Unbound subhaloes may appear to be part of a cluster in projection, and this can bias velocity dispersion measurements. However, the same member selection and velocity dispersion algorithms used here were applied to mock catalogs including this 'interloper' population in \cite{old15}, who showed that despite this our method is able to recover unbiased mass measurements. Instead, we aim to assess any intrinsic, \emph{relative} biases introduced in our sample by having different observing strategies. We account for the impact of these projection effects as a systematic uncertainty in our final estimate of the dynamical mass uncertainties (see \Cref{s:mdyn}).

To simulate the southern observations, we observed up to 70 galaxies in the inner 
$5\arcmin\times5\arcmin$ randomly. Given the resolution of Multidark, we were typically able to 
``observe'' 45 subhaloes following this strategy. To recreate the equatorial observations we first 
identified the approximate major axis of the subhalo distribution for each cluster by taking the 
mean direction of the 10 largest distances between cluster members. We then created a MOS mask 
with slits of length $8\arcsec$ along this major axis (axis $x'$ as per the right panel of 
\Cref{f:observations}), and observed exactly one subhalo in each slit (unless there were no 
subhaloes in the slit area). To define which subhalo to ``observe,'' we selected an object from 
each slit with a Gaussian distribution around the cluster major axis such that we preferentially, 
but not exclusively, observed subhaloes close to the line passing through the central subhalo 
(representing the BCG). This setup led to, on average, 25 subhaloes observed per cluster for the 
equatorial strategy. \Cref{f:observations} illustrates our southern and equatorial spectroscopic 
strategies applied to a halo of the Multidark simulation. Because the number of ``observed'' 
subhaloes per halo is lower than the number of observed galaxies per cluster, the statistical 
uncertainties in the velocity dispersions from the mock observations overestimate the measured 
uncertainties per cluster. This, however, does not compromise our assessment of a bias introduced 
by either strategy, and is compensated by the large number of simulated haloes used. We note that 
the strategies defined above are generalizations (e.g., some clusters in the equator have denser 
sampling and out to smaller radii). We apply the relevant corrections (see below) to all clusters 
irrespective of the sample they belong to (that is, southern or equatorial sample), solely based 
on their particular observational setup.

\begin{figure}
 \centerline{\includegraphics[width=3.4in]{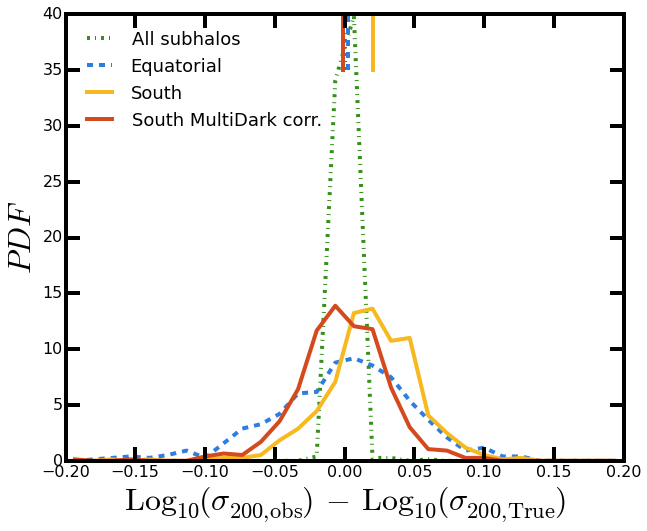}}
\caption{Probability distribution functions of the residuals between the measured and true 
velocity dispersions in the Multidark simulation, in logarithmic space. The dash-dotted green, 
dashed blue, and solid yellow lines show the differences using all subhaloes, an average of 25 
subhaloes with the equatorial strategy, and an average of 45 subhaloes with the southern strategy, 
respectively. The red solid line shows the residuals in the southern strategy after correcting for 
incomplete sky coverage (see \Cref{s:sims}). Vertical lines at the top of the figure show the 
median values.}
\label{f:veldisp_sims}
\end{figure}

The residuals in the recovered velocity dispersions with respect to the true halo velocity 
dispersion (i.e., that determined using all subhaloes, typically 60) are shown in 
\Cref{f:veldisp_sims} for each observational strategy. As a consistency check, we also show the 
residuals determined from measuring the velocity dispersion from all subhaloes within $r_{200}$, as 
determined iteratively from the mock observations following the procedure described in 
\Cref{s:vdisp}, which are consistent with the true velocity dispersions within the statistical 
uncertainty. This comparison shows that the adopted scaling relation (see \Cref{s:scaling}) is 
consistent with the scaling of Multidark haloes and that, in an ideal case where we observe all 
subhaloes, our estimates of both $\sigma_{200}$ and $r_{200}$ (and thereby $M_{200}$) are unbiased.

\begin{figure}
 \centerline{\includegraphics[width=3.4in]{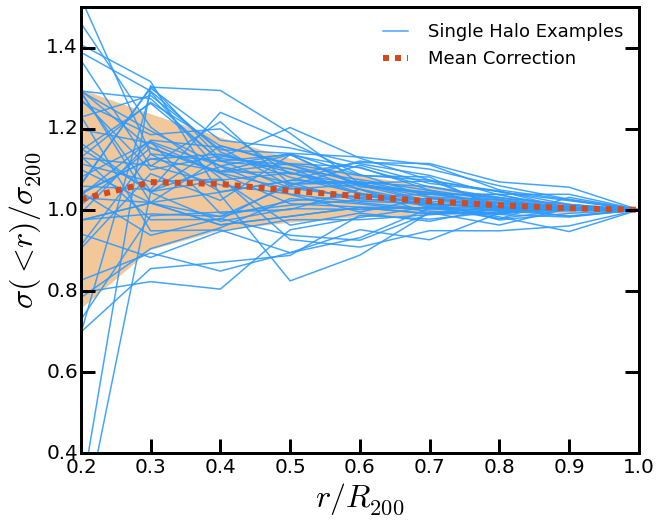}}
\caption{Enclosed one-dimensional velocity dispersion, $\sigma(<r)$, as a function of radius, $r$, for subhaloes in Multidark, normalized to $\sigma_{200}$ and $r_{200}$, respectively. The red dashed line is the mean value, and the orange region encloses 68 per cent of the haloes. Blue lines are a random subset of the Multidark haloes. The data for this figure are presented in \Cref{t:MDcorr}.}
\label{f:correction}
\end{figure}

\Cref{f:veldisp_sims} also shows the distribution of velocity dispersions recovered from the simulations for both our observing strategies. For the equatorial strategy the velocity dispersions are unbiased, meaning that the velocity distributions are well sampled within the statistical precision we require---there is no bias introduced by sampling galaxies along a particular direction \cite[but see][ for evidence of a preferred direction for the velocity distribution in galaxy clusters]{skielboe12}. The velocity distribution derived from the southern strategy is, on the other hand, biased by 0.02 dex (corresponding to $\approx5$ per cent) on average, which is consistent with the picture of a decreasing velocity dispersion outward from the cluster centre \citep[e.g.,][]{mamon10}. We correct for this bias by measuring the \textit{true} integrated velocity dispersion profiles for Multidark haloes and scaling them up to $\sigma_{200} \equiv \sigma(r=r_{200})$. We list the radial correction $\sigma(<r)/\sigma_{200}$ and the associated scatter in \Cref{t:MDcorr}, and show it in \Cref{f:correction}. While the spread increases towards small apertures, the correction is $<10$ per cent at all radii. However, at small radii the scatter is large and must be included in the error estimate when measuring velocity dispersions. We apply this correction to each halo when observed with the southern strategy and are able to recover unbiased velocity dispersions (see \Cref{f:veldisp_sims}).

\begin{table}
\centering
\caption{Ratio of one-dimensional velocity dispersion within an aperture $r$, $\sigma(<r)$, to the one-dimensional velocity dispersion within $r_{200}$, $\sigma_{200}$, estimated using subhaloes in the Multidark simulation. Uncertainties are the standard deviations. These values are plotted in \Cref{f:correction}.}
\label{t:MDcorr}
\begin{tabular}{c c}
\hline\hline
$r/r_{200}$ & $\langle\sigma(<r)/\sigma_{200}\rangle$ \\[0.5ex]
\hline
0.2 & $1.03\pm0.27$ \\
0.3 & $1.07\pm0.17$ \\
0.4 & $1.06\pm0.11$ \\
0.5 & $1.05\pm0.08$ \\
0.6 & $1.03\pm0.05$ \\
0.7 & $1.02\pm0.04$ \\
0.8 & $1.01\pm0.02$ \\
0.9 & $1.00\pm0.01$ \\
1.0 & $1.00\pm0.00$ \\
\hline
\end{tabular}
\end{table}

\subsection{From velocity dispersions to dynamical masses}
\label{s:scaling}

\begin{figure}
 \centerline{\includegraphics[width=3.4in]{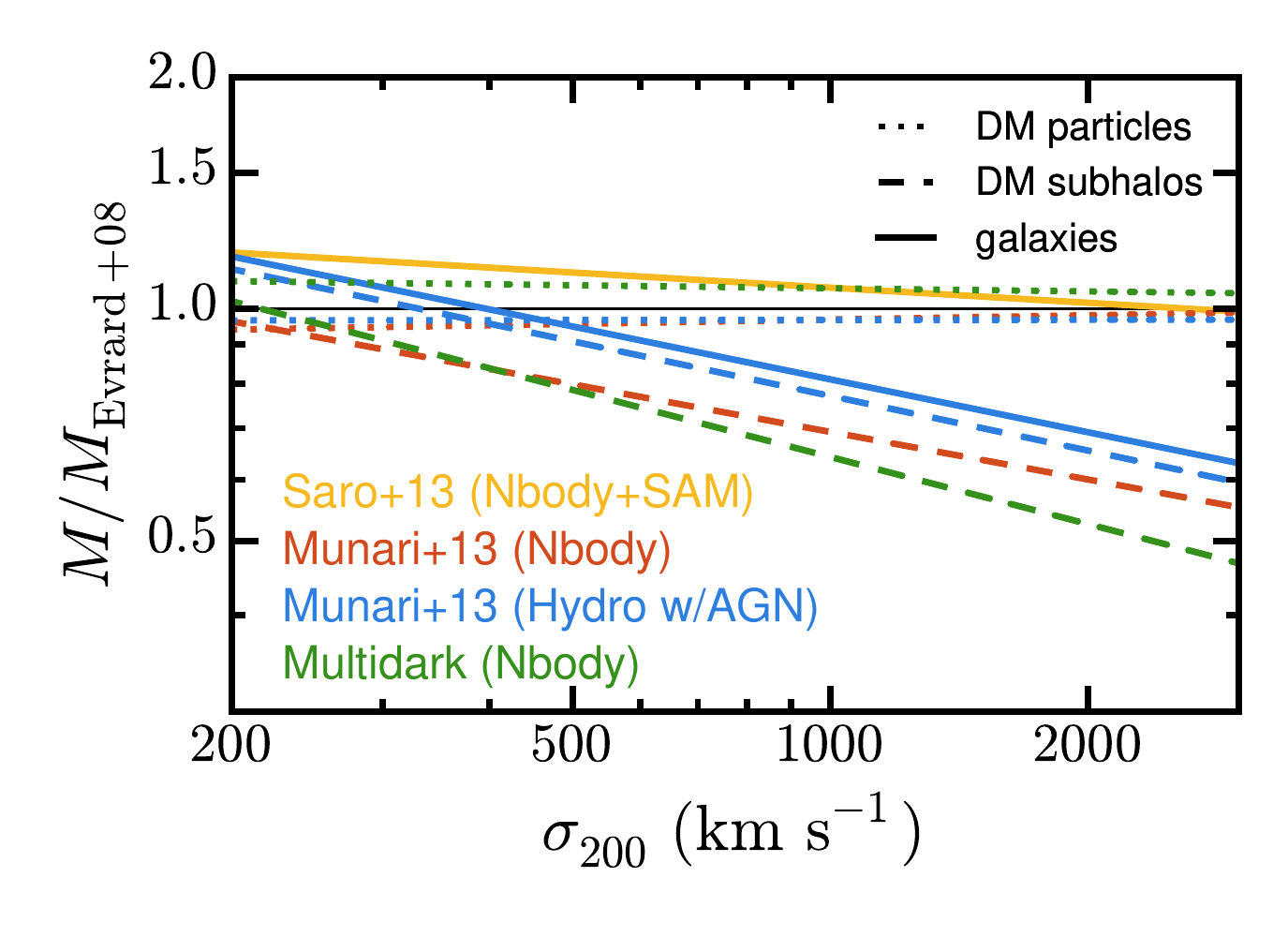}}
\caption{Comparison between different $\sigma-M$ scaling relations, relative to the scaling 
relation of dark matter particles derived by \citet{evrard08}. Dotted, dashed and solid lines show 
scaling relations for dark matter particles, dark matter subhaloes and galaxies, respectively. We 
show the scaling relation for galaxies derived from a semi-analytic model implemented in a dark 
matter-only simulation by \citet{saro13} in yellow. Red and blue lines show the scaling relations 
derived from dark matter-only and full hydrodynamical simulations, respectively, by 
\citet{munari13}, and the green lines show scaling relations from the Multidark simulation. In 
this work, we calculate dynamical masses using the scaling relation given by the blue solid line.}
\label{f:scalings_simulations}
\end{figure}
 
In \cite{sifon13}, we used the $\sigma-M_{200}$ scaling relation of \cite{evrard08} to estimate 
dynamical masses. As discussed in \Cref{s:intro}, the scaling relation of \cite{evrard08} was 
calibrated from a suite of \Nbody\ simulations using dark matter particles to estimate velocity 
dispersions. However, the galaxies, from which velocity measurements are made in reality do not 
sample the same velocity distribution as the dark matter (hereafter DM) particles. They feel 
dynamical friction and some are tidally disrupted, which distorts their velocity distribution and 
biases their dispersion \citep[e.g.,][]{carlberg94,colin00}. Recent high resolution hydrodynamical 
simulations of ``zoomed'' cosmological haloes have shown that there is a significant difference 
between the velocity distributions of DM particles and galaxies themselves; whether galaxies 
(i.e., overdensities of stars in hydrodynamical simulations) or dark matter subhaloes are used 
makes comparatively little difference \citep{munari13}. Results from state-of-the art numerical 
simulations depend on the exact definition of a galaxy and the member selection applied, but the 
current consensus is that galaxies are biased high (i.e., at a given mass the velocity dispersion 
of galaxies or subhaloes is larger than that of DM particles) by 5$-$10 per cent with respect to DM 
particles \citep{lau10,munari13,wu13}, translating into a positive 15$-$20 per center bias in 
dynamical masses when using DM particles. This is illustrated in \Cref{f:scalings_simulations}: DM 
particles are not significantly impacted by either dynamical friction or baryonic physics; 
therefore the scaling relations for DM particles are essentially the same for all simulations. In 
contrast, dark matter subhaloes are affected by baryons in such a way that including baryonic 
feedback (most importantly feedback from active galactic nuclei (AGN), but also from cooling and 
star formation) makes their velocity dispersions much more similar to those of simulated galaxies. 
This means we can rely on our analysis of the previous section, based on dark matter subhaloes, to 
correct the velocity dispersions measured for ACT clusters, and then estimate dynamical masses 
using predictions obtained either from galaxies or subhaloes. The difference between the 
\cite{saro13} and \cite{munari13} galaxy scaling relations depends on the details of the 
semi-analytic and hydrodynamical implementations used in \cite{saro13} and \cite{munari13}, 
respectively. The different cosmologies used in the Millenium simulation \citep[in particular, 
$\sigma_8=0.9$;][]{springel05_millenium} by \cite{saro13} and the simulations by \cite{munari13} 
($\sigma_8=0.8$) may also play a role.

We therefore use the scaling relation between the projected galaxy velocity dispersion and mass 
estimated by \cite{munari13}, obtained from zoomed-in hydrodynamical simulations of dark matter 
haloes, that includes prescriptions for cooling, star formation, and AGN feedback,
\begin{equation}\label{eq:munari}
 \sigma_{\rm 200} = A_{\rm 1D} \left[\frac{h\,E(z)\,M_{200}}{10^{15}\Msun}\right]^\alpha
\end{equation}
where $\sigma_{200}$ is the 3-dimensional velocity dispersion of galaxies within $r_{200}$, 
divided by $\sqrt{3}$ (i.e., the line-of-sight velocity dispersion in a spherical cluster), 
$E(z)=[\Omega_\Lambda + (1+z)^3\Omega_{\rm m}]^{1/2}$, $A_{\rm 
1D}=1177\pm4.2\,\kms$, and $\alpha=0.364\pm0.002$. The intrinsic scatter \emph{at 
fixed mass} in \Cref{eq:munari} is of order 5 per cent, or $\approx$15 per cent in mass 
\citep{munari13}, but this value does not include the effect of interlopers (that is, impurity in 
the member sample), which can increase the intrinsic scatter by up to a factor two 
\citep{biviano06,mamon10,white10,saro13}. This is an irreducible uncertainty since there is always a 
fraction of contaminating galaxies that cannot be identified by their peculiar velocities because 
they overlap with the velocity distribuion of actual members \citep[see, e.g., figure 10 
of][]{white10}. Hence we adopt a figure of 30 per cent for each cluster's mass uncertainty arising from interlopers in the member sample. Note that we automatically account for the velocity bias, $b_{\rm v}$, by adopting a scaling relation based on simulated galaxies rather than dark matter particles (see \Cref{s:mdyn} for further discussion).

The velocity dispersion measurements were obtained for a pre-selected set of clusters, and the 
sample was not further refined based on these measurements. So although the measurements are 
affected by noise and intrinsic scatter, we can expect positive and negative noise and scatter 
excursions to be equally likely. The dynamical mass measurements on this sample are thus not 
affected by Eddington bias; this is discussed further in \Cref{ap:bias}. We therefore
calculate dynamical masses by directly inverting \Cref{eq:munari}, which gives $\sigma(M)$, in 
order to obtain $M(\sigma)$. For this computation we take the uncertainty on $\sigma$ to be 
normal, and report the mean and standard deviation of $M(\sigma)$ after propagating the full error 
distribution.\footnote{The error distribution is normal in $\sigma$ but not in 
$M\propto\sigma^{1/\alpha}$ (with $1/\alpha\approx3$). Therefore the mean mass is not the cube of 
the central value of $\sigma$. This difference depends only on the measurement uncertainty and for 
our sample its median is 3 per cent, with a maximum of 16 per cent for ACT-CL~J0206.2$-$0114.} We note that this procedure can yield biased dynamical masses if velocity dispersion and SZ effect measurements are correlated for individual clusters \citep{evrard14}. In fact, we may expect some degree of correlation between any pair of observables for a given cluster, because the same large scale structure is affecting all cluster observables \citep{white10}. We defer a proper treatment of correlations between observables to future work.

\subsection{Dynamical mass estimates}
\label{s:mdyn}

\begin{table}
\centering
\caption{Individual cluster mass uncertainty budget, given as a fraction of cluster mass. Central 
values are the medians of the cluster distributions and uncertainties are 16th and 84th 
percentiles; upper limits are 84th percentiles. The median is equal to zero for all values with 
upper limits. ``Reported'' uncertainties correspond to those in \Cref{t:masses}, which arise from the combination of the three effects preceding them, while ``total'' uncertainties include the 30 per cent scatter from the $\sigma-M$ scaling relation, which is fixed for all clusters, added in quadrature. The 15\percent\ uncertainty in the velocity bias is an overall uncertainty on the average masses.}
\label{t:uncertainties}
\begin{tabular}{l c c c}
\hline\hline
 Source & Equator & South & All \\[0.5ex]
\hline
 Statistical & $0.31_{-0.08}^{+0.21}$ & $0.25_{-0.03}^{+0.09}$ & $0.28_{-0.06}^{+0.20}$ \\[0.3ex]
 Member selection & $0.14_{-0.14}^{+0.18}$ & $<0.01$ & $0.04_{-0.04}^{+0.18}$ \\[0.3ex]
 Multidark correction & $<0.12$ & $0.07_{-0.07}^{+0.18}$ & $<0.18$ \\[0.2ex]
 \hline
 {\bf Reported} & {\bf 0.36} & {\bf 0.26} & {\bf 0.31} \\[0.2ex]
\hline
 Scatter in $M(\sigma)$ & 0.30 & 0.30 & 0.30 \\[0.2ex]
\hline
 {\bf Total} & {\bf 0.47} & {\bf 0.40} & {\bf 0.44} \\[0.2ex]
\hline
 Velocity bias uncertainty & 0.15 & 0.15 & 0.15 \\[0.2ex]
\hline
\end{tabular}
\end{table}

The masses thus estimated are listed in \Cref{t:masses}, along with the redshifts, velocity 
dispersions, number of members used and $r_{200}$. We also list the radius at which our 
spectroscopic coverage ends, $r_{\rm max}$, and the initial velocity dispersion measured within 
$r_{\rm max}$. Below we summarize the corrections applied with respect to our analysis in 
\cite{sifon13} and then present a detailed account of uncertainties entering our dynamical mass 
estimates, before comparing our mass estimates with masses derived from SZ measurements.

Two sources of bias are now accounted for that were not included in \cite{sifon13}. The first is 
the radial coverage of spectroscopic members \citep[which was discussed, but not corrected for, 
in][]{sifon13} which includes (i) an iterative calculation of the velocity dispersion within 
$r_{200}$ only for 24 clusters with $r_{\rm max}>r_{200}$ and (ii) a correction to the velocity 
dispersion, based on the velocity dispersion profile of subhaloes in the Multidark simulation (see 
\Cref{f:correction}), for 20 clusters with $r_{\rm max}<r_{200}$. Over the full sample these two 
situations produce a net correction of $-$5 per cent, compared to applying no correction as in 
\cite{sifon13}. The second source of bias is the relation between the velocity dispersion of dark 
matter particles and that of galaxies. We account for this difference by using the $\sigma-M$ 
scaling relation of galaxies derived by \cite{munari13}, which gives average masses 20 per cent 
lower than those derived from the scaling relation of \cite{evrard08} used in \cite{sifon13}. In 
addition,  we have updated the minimum bin size in our member selection algorithm (cf.\ 
\Cref{s:vdisp}), which lowers the masses by an average 6 per cent with respect to the value 
adopted in \cite{sifon13}. Because of these updates to our analysis, for the southern clusters we 
report masses that are, on average, $(71\pm8)$ per cent of those reported in \cite{sifon13}. 

We present a breakdown of the contributions to individual cluster mass uncertainties for the equatorial and southern samples in \Cref{t:uncertainties}. The uncertainty budget is dominated by the scatter induced by interlopers. Based on the discussion presented in \Cref{s:scaling}, we estimate that this uncertainty amounts roughly to 30\percent\ in mass. Because this contribution corresponds to a constant uncertainty for all clusters, we do not include them in the uncertainties reported in \Cref{t:masses}. We do recommend this 30\percent\ systematic (i.e., that cannot be reduced by observing more galaxies) uncertainty to be added to the reported uncertainties in any cosmological analysis that uses these dynamical masses, and we include it in our calculation of the SZ mass bias in \Cref{s:msz}.
Similarly, based on the discussion in \Cref{s:intro}, we adopt a 15 per cent systematic uncertainty arising from the unknown velocity bias \citep[5\percent\ in velocity; e.g.,][]{wu13}. This 15 per cent essentially accounts for i) the fact that our galaxy sample may not correspond to the galaxy samples used by \cite{munari13} to arrive at \Cref{eq:munari} (because the velocity bias is luminosity-dependent), and ii) differences in scaling relations compared to that of \cite{munari13} that may arise because of different hydrodynamical implementations (each producing a different velocity bias). The unknown velocity bias therefore limits our constraints on the SZ mass bias (see \Cref{s:msz}).

Statistical uncertainties are the dominant contribution to the reported uncertainties, with a median contribution of 28\percent\ of the cluster mass. Uncertainties from member selection and the scatter in the correction of \Cref{t:MDcorr} are subdominant. We note here that by ``member selection'' we mean uncertainties arising from including or rejecting particular galaxies through the shifting gapper. The true uncertainty from contaminating galaxies is included in the scaling relation scatter as discussed in \Cref{s:scaling}. We make this distinction because there is a fraction of false members which cannot be identified observationally via their peculiar velocities \citep[e.g.,][]{mamon10,white10,saro13}. For the SZ-selected clusters of the South Pole Telescope survey, \cite{ruel14} found that the uncertainty from member selection, estimated by ``pseudo-observing'' their stacked cluster, depends on the number of galaxies observed. For the number of members we observed \citep[which is roughly a factor two larger than the average number of members observed by][]{ruel14}, their estimate of the combined statistical and member selection uncertainty is consistent with ours.

\subsection{Comparison to SZ-derived masses}
\label{s:msz}

The usefulness of clusters for constraining cosmological parameters depends on the accurate 
calibration of the cluster mass scale. Calibrated SZ masses are especially informative because SZ 
surveys yield large samples of clusters reaching to high redshifts. While our dynamical and SZ 
mass proxies may have non-trivial mass or redshift-dependence, the data in our study permit us to
constrain the average bias between these proxies within the mass range probed in this study.

\begin{figure}
 \centerline{\includegraphics[width=3.4in]{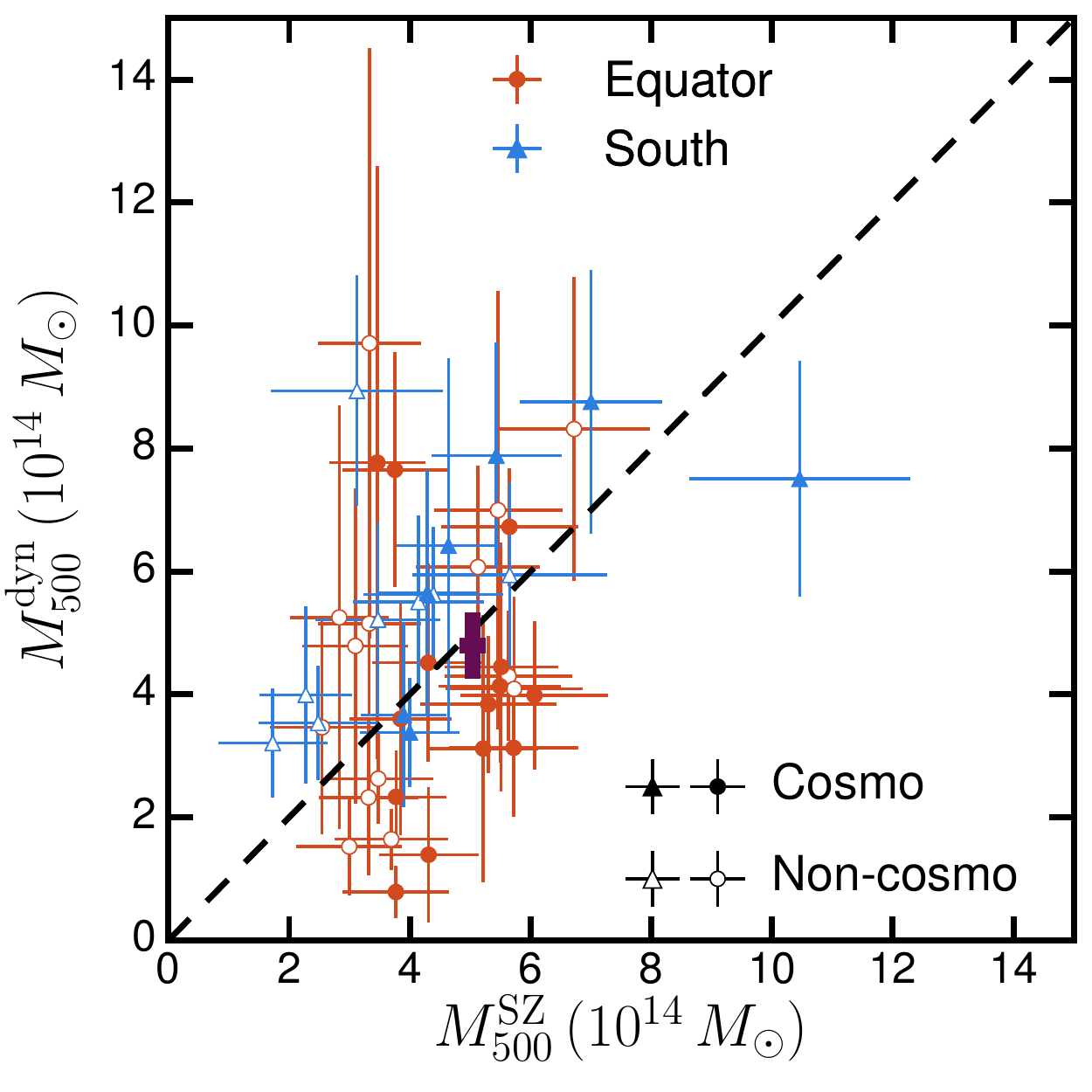}}
\caption{Comparison of dynamical and SZ-derived masses. Red circles and blue triangles show clusters in the equatorial and southern samples, respectively. Uncertainties on dynamical masses correspond to those reported in \Cref{t:masses}, and for clarity do not include the 30\percent\ scatter in the $M(\sigma)$ relation. The dashed line shows equality and the purple cross shows the average SZ and dynamical masses of the combined southern and equatorial cosmological samples (filled symbols), calculated including the 30\percent\ intrinsic scatter in the $M(\sigma)$ relation on the dynamical masses. There is an additional 15\percent\ overall uncertainty on dynamical masses arising from the unknown galaxy velocity bias. See \Cref{s:msz} for details.}
\label{f:msz}
\end{figure}

We compare the dynamical masses to the SZ-derived masses, $M_{500}^{\rm SZ}$, in \Cref{f:msz}. For 
the purpose of this comparison we rescale dynamical masses to $M_{500}$ using the 
mass-concentration relation of \cite{dutton14}. The SZ-derived masses assume a scaling relation 
between the SZ effect (specifically, $Y_{500}$) and mass based on the pressure profile of 
\cite{arnaud10}, derived from X-ray observations of local ($z<0.2$) clusters, and have been 
corrected for Eddington bias as detailed in \cite{hasselfield13}, assuming a 20 per cent intrinsic 
scatter in $Y_{\rm 500}$ at fixed true mass \citep[the ``UPP'' masses of][]{hasselfield13}. We 
refrain from fitting a scaling relation to these data since this requires a proper calibration of 
the survey selection effects and accounting for the mass function and cosmological parameters; the 
dynamical mass--SZ scaling relation and inferred cosmological parameters will be presented in a 
future paper.

Beyond the assumptions used to obtain the SZ masses, any additional bias in the inferred mass 
relative to the true cluster mass is often parametrized in terms of the SZ mass bias, $1-b_{\rm 
SZ}$ \citep[e.g.,][]{planck15xxiv}, defined by the relation $\langle M_{\rm SZ} \vert M_{\rm true} 
\rangle = (1-b_{\rm SZ})M_{\rm true}$. An understanding of this calibration is essential to the 
cosmological interpretation of cluster counts from SZ surveys. Similarly, our dynamical masses may 
be biased proxies for the true cluster mass. Following \cite{hasselfield13}, we parametrize this 
bias with $\beta_{\rm dyn}$, defined by $\langle M_{\rm dyn} \vert M_{\rm true} \rangle \equiv 
\beta_{\rm dyn} M_{\rm true}$. For the remainder of this section we use the word ``bias'' to refer 
to systematic effects on the sample, such as ``Eddington bias,'' that do not average down to an 
expectation value of zero with an increasing sample size.

The SZ and dynamical mass data permit us to place limits on the ratio $(1-b_{\rm SZ})/\beta_{\rm 
dyn}$ by comparing the average SZ and dynamical masses of the clusters from the cosmological 
sample. We first combine the dynamical masses into a single characteristic mass
\begin{align}
\bar{M}_{\rm dyn} \equiv \frac{\sum_i w_i M_{i,{\rm dyn}}}{\sum_i w_i},
\end{align}
where the $M_{i,{\rm dyn}}$ represent the individual dynamical mass measurements, and the $w_i$ 
are weighting factors. For this analysis we set all $w_i = 1$, but see below for further 
discussion of the choice of weights. We also compute the error, through standard error 
analysis.\footnote{The error in $\bar{M}$ is $(\sum_i w_i^2 \epsilon_i^2)^{1/2}/(\sum_i w_i)$, where 
$\epsilon_i$  is the error in $M_i$.} For the SZ masses, we form the analagous sum, $\bar{M}_{\rm 
SZ} \equiv (\sum_i w_i M_{i,{\rm SZ}})/(\sum_i w_i)$.  Note that the weights $w_i$ used
in this expression are the same weights used to compute $\bar{M}_{\rm dyn}$; this is essential to 
obtaining an unbiased answer when we later combine the two characteristic masses.

Each mass measurement is contaminated by intrinsic scatter and noise, in the sense that
\begin{align}
  \bar{M}_{i,{\rm dyn}} & = M_{i,{\rm true}}e^{\xi_{i}} + \delta M_{i},
\end{align}
where $\delta M_{i,{\rm true}} \sim \mathcal{N}(0,\epsilon_i)$ is the contribution from measurement 
noise, and $\xi_i \sim \mathcal{N}(0,0.3)$ is the contribution from intrinsic scatter. The 
expectation value for $\delta M_{i}$ is zero, while the expectation value of $e^{\xi_{i}}$ is 
1.046. So when we combine our measurements into a characteristic mass we expect that
\begin{align}
 \langle \bar{M}_{\rm dyn} \rangle & = \frac{\sum_i w_i \langle M_{i,{\rm dyn}} \rangle}{\sum_i 
   w_i} =  1.046\,\beta_{\rm dyn}\frac{\sum_i w_iM_{i,{\rm true}}}{\sum_i w_i}.
\end{align}

For the combination of the SZ masses, the expectation value is $\langle \bar{M}_{\rm SZ} 
\rangle=1-b_{\rm SZ}$, because the skewness introduced by intrinsic scatter has already been fully 
accounted for in the calculation of the $M_{\rm SZ}$ values used here by \cite{hasselfield13}.  
Taking the ratio of these two characteristic masses gives
\begin{align}
  \frac{\langle \bar{M}_{\rm SZ} \rangle}{\langle \bar{M}_{\rm dyn} \rangle} &
   = \frac{(1-b_{\rm SZ})}{1.046\,\beta_{\rm dyn}}.
\end{align}

Our measured values of $\bar{M}_{\rm SZ}$ and $\bar{M}_{\rm dyn}$ thus provide a useful meaurement 
of $(1-b_{\rm SZ}) / \beta_{\rm dyn}$. (We show in \Cref{ap:bias} that this ratio is unbiased.) 
For the 21 clusters in the cosmological sample, the characteristic dynamical mass under uniform 
weights is $\bar{M}_{\rm dyn} = (4.8 \pm 0.5) \times 10^{14} \Msun$, and the characteristic SZ 
mass is $\bar{M}_{\rm SZ} = (5.0 \pm 0.2) \times 10^{14} \Msun$. The ratio of calibration factors 
is then
\begin{align}\label{eq:massbias}
  \frac{(1-b_{\rm SZ})}{\beta_{\rm dyn}} = \massbias \, ({\rm stat.}) \, \pm 0.14 \, ({\rm syst.}) \,
\end{align}
where the 0.14 systematic uncertainty arises from the 15\percent\ fractional uncertainty on the average dynamical masses due to the unknown velocity bias discussed in \Cref{s:mdyn}. We note again that, in computing \Cref{eq:massbias}, we have accounted for the 30 per cent scatter in the $M(\sigma)$ relation (see \Cref{s:scaling}) which is not included in the cluster mass uncertainties reported in \Cref{t:masses} and shown in \Cref{f:msz}. Recent estimates of the SZ mass bias combining weak lensing measurements and SZ mass estimates from \cite{planck15xxvii} have found $M_{\rm SZ}/M_{\rm WL}\approx0.7$ \citep{vonderlinden14_planck,hoekstra15}. These measurements were then used as priors for the cosmological analysis of Planck SZ-selected clusters \citep[under the assumption that $\langle{M}_{\rm WL}\rangle=\langle{M}_{\rm true}\rangle$,][]{planck15xxiv}, highlighting the importance of calibrating these biases. We note that both \cite{vonderlinden14_planck} and \cite{hoekstra15} probed higher masses than we do. In fact, both works found evidence (at 95 per cent confidence) for a mass-dependent bias which, at the typical masses of ACT clusters, is consistent with our estimate. Similar to us, \cite{rines16} found no evidence that the mass ratio $\langle\bar{M}_{\rm SZ}\rangle/\langle\bar{M}_{\rm dyn}\rangle$ is different from unity, using  dynamical masses estimated with the caustic technique, in a mass regime similar to ours.

\cite{battaglia16} used a stacked weak lensing measurement on a subset of these clusters, which 
they fit using hydrodynamical simulations, and found an SZ mass bias $1-b_{\rm 
SZ}=\bar M_{\rm SZ}/\bar M_{\rm WL}=0.98\pm0.28$ (assuming $\langle M_{\rm WL} \vert M_{\rm 
true} \rangle = M_{\rm true}$, as has been assumed in recent studies). This value has been 
computed with weights that depend on the weak lensing measurements. As a consistency check, we 
estimate the average dynamical mass of the nine clusters used by \cite{battaglia16}, using the 
same weak lensing weights, and find $\bar M_{\rm dyn-WL}=(4.7\pm1.4)\times10^{14}\Msun$, which 
implies a mass ratio $\bar M_{\rm dyn-WL}/\bar M_{\rm WL}=0.98\pm0.33$. Therefore the dynamical 
masses are consistent with the weak lensing masses derived by \cite{battaglia16}.

We have used uniform weights $w_i=1$ to obtain the ratio in \Cref{eq:massbias}, but one might 
expect that more carefully chosen weights could provide a more precise answer. In fact the weights 
should be chosen with some care, as it is possible to introduce a bias into this ratio if one 
permits the weights to depend too much on the measured data themselves. For example, if we take 
weights $w_i = 1/\epsilon_{i,{\rm dyn}}^2$ (where $\epsilon_{i,{\rm dyn}}$ is the measurement uncertainty on the velocity dispersion of cluster $i$), we find that clusters with low dynamical mass are more 
strongly weighted, because the dynamical mass uncertainties are strongly correlated with the 
dynamical mass measurements. However, the SZ masses are limited by sample selection effects to lie 
above some minimum value, and the characteristic SZ mass under these weights is almost twice the 
characteristic dynamical mass. A somewhat weaker effect is that these weights (i.e., the $w_i$ 
above) have the potential to introduce a sort of Eddington bias into the computation, even though 
we carefully constructed a sample for which the dynamical mass measurements were unbiased. By 
re-weighting the clusters in a way that is correlated with the dynamical mass measurements 
themselves, we are effectively sub-selecting for measurements that are more likely to have  
scattered below the true mass values.

To avoid such biases, one should not incorporate dynamical mass or its uncertainty into the 
weights. Similarly, one should be wary of using the measured SZ masses and uncertainties to set 
the weights. In the present data, using weights $w_i = 1/\epsilon_{i,{\rm SZ}}^2$ or $w_i =
M_{i,{\rm SZ}}^2/\epsilon_{i,{\rm SZ}}^2$ changes the resulting ratio by +3 or -2 per cent, 
respectively, without reducing the uncertainty. We discuss alternative weighting schemes in 
\Cref{ap:bias}.

\subsection{Cluster substructure}\label{s:substruct}

Probes of substructure within a galaxy cluster provide information on a cluster's dynamical state: 
for example, whether or not a recent merger event has occurred. Since the thermodynamic properties 
of clusters vary depending on their dynamical state, a measurement of the amount of substructure 
provides an important additional cluster property. Of particular interest to SZ experiments is how 
the integrated $Y$ parameter fluctuates with dynamical state. Simulations have shown that $Y$ can 
fluctuate by tens of percent shortly after merger events 
\cite[e.g.,][]{poole07,wik08,krause12,nelson12}, and that the intrinsic scatter of the $Y-M$ 
scaling relation for a subsample of relaxed clusters is smaller than for the entire sample 
\citep[e.g.,][]{battaglia12,yu15}. The latter conclusion was also reached by \cite{sifon13}, 
albeit with low statistical significance.

Because of the sparser spectroscopic sampling used for the equatorial clusters, it is more 
difficult to identify localized substructure than it is with our dense sampling of the southern 
clusters (which is, however, confined to a smaller region of the cluster). We therefore refrain 
from a detailed, cluster-by-cluster analysis of substructure in the equatorial sample as we did 
for the southern clusters in \cite{sifon13}. However, it is still valuable to study the presence 
of substructure in the sample as a whole, to be able to compare between our equatorial and 
southern samples, and whether the SZ selects different cluster populations than other techniques.

We use two quantities used by \cite{sifon13} to study cluster substructure. The first is the 
peculiar velocity of the BCG, $\vbcg=c(z_{\rm BCG}-z_{\rm cl})/(1+z_{\rm cl})$, where $z_{\rm cl}$ 
is the cluster redshift listed in \Cref{t:masses}. Based on the results of 
\Cref{s:compare_measurements}, we assume an error of $100\,\kms$ on $\vbcg$. The 
second estimator we use is the DS test \citep{dressler88}, which measures the deviation of the 
velocity distribution in localized regions of a cluster with respect to the cluster as a whole 
through the statistic $\Delta=\sum_i\delta_i$, where
\begin{equation}\label{eq:delta}
\delta^2 = \frac{N_{\rm local}}{\sigma^2}\left[\left(\bar v_{\rm local} - \bar v\right)^2 + 
\left(\sigma_{\rm local} - \sigma\right)^2\right]^2
\end{equation}
is calculated for each galaxy, where $\bar v_{\rm local}$ and $\sigma_{\rm local}$ are the mean 
and dispersion of the velocity distribution of the $N_{\rm local}$ nearest neighbors, where 
typically $N_{\rm local}=\sqrt{N_{200}}$. For each cluster we compare $\Delta$ to 1000 
realizations where we shuffle the galaxy velocities, keeping their positions fixed. $\sdelta$ is 
then the probabilitiy to exceed the $\Delta$ measured for the clusters, given statistical 
fluctuations as determined through these realizations. We calculate 68 per cent level 
uncertainties on $\sdelta$ by varying $N_{\rm local}$ in the range $\sqrt{N_{200}}-3 \leq N_{\rm 
local} \leq \sqrt{N_{200}}+3$. Typically, $\sdelta<0.05$ is taken as evidence for substructure 
\citep{pinkney96}. See \cite{sifon13} for more details.

\begin{figure}
 \centerline{\includegraphics[width=3.2in]{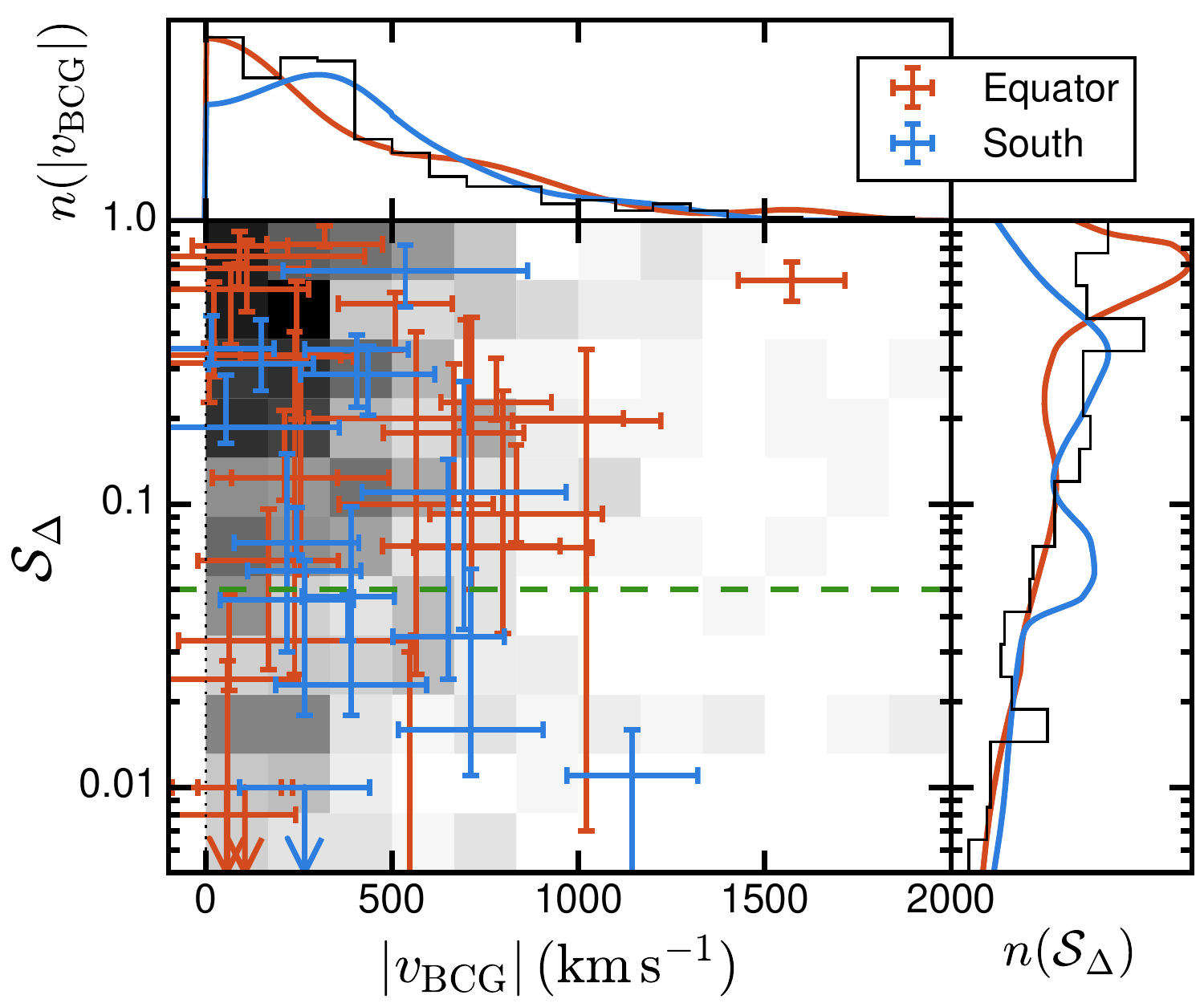}}
\caption{Comparison between BCG peculiar velocity and DS test significance level for the southern 
(blue) and equatorial (red) samples, and haloes in the Multidark simulation (grey scale background, where zero is white and one is black, and black histograms). The green, dashed horizontal line shows the threshold $\sdelta=0.05$, below which the DS test is usually considered to provide evidence for substructure. The \textit{top} and \textit{right} panels show the corresponding histogram for the Multidark simulation and the summed probability distribution functions for the southern (blue) and equatorial (red) clusters, normalized to subtend the same area as the histograms. 
}
\label{f:substructure}
\end{figure}

We compare in \Cref{f:substructure} the distributions of (absolute values of) BCG peculiar 
velocities, $\vbcgabs$, and $\sdelta$ to those found in the Multidark simulation. In general, the 
southern sample shows more evidence of substructure than the equatorial sample through the DS test, 
with 38 per cent and 22 per cent, respectively, having $\sdelta<0.05$. In turn, 31 per cent of 
Multidark clusters fulfill this criterion. Two-sided Kolmogorov-Smirnov tests, however, show no evidence of the distributions of either $\sdelta$ or $\vbcgabs$ being different between the southern, equatorial and Multidark samples (all $p$-values from the KS tests are $\gtrsim0.20$). In \cite{sifon13}, we selected clusters as non-relaxed (i.e., containing substructure) if, among other properties, they had $\vbcgabs>0$ at 95 per cent confidence. In the southern sample 50 per cent of clusters pass this test, while 41 per cent (11/27) of the clusters in the equatorial sample do. This 40$-$50 per cent rate of non-relaxed clusters is somewhat lower than fractions found for X-ray-- and optically--selected clusters \citep[e.g.,][]{bohringer10,wen13}. Using mock cluster observations, \cite{lin15} showed that the SZ significance can be boosted by up to 10 per cent for cool core (i.e., relaxed) clusters depending on the redshift, cuspiness and size of the cluster. This would then lead to a preferential selection of relaxed clusters, qualitatively consistent with our results. The fact that this bias is not apparent when comparing to Multidark clusters may relate to the fact that Multidark is a dark matter only simulation, but a detailed comparison is beyond the scope of this work.

\subsection{The impact of centring on the BCG}
\label{s:miscentring}

As mentioned in \Cref{s:vdisp}, we have assumed that the BCGs correspond to the centre of the 
cluster potential. In this section we estimate the impact of this assumption on the reported 
masses.

We first re-calculate the velocity dispersions for all clusters assuming that the cluster centre 
is the centre of light instead of the BCG. To estimate the centre of light we take the 
luminosity-weighted average position of photometric members using photometric redshifts estimated 
by \cite{menanteau13} using the Bayesian Photometric Redshift (BPZ) code \citep{benitez00}. The 
average mass ratio is $\langle M_{\rm CoL}/M_{\rm BCG} \rangle = 1.01\pm0.09$ with a standard 
deviation of 0.16, which is within the quoted mass uncertainties. Note that \cite{viola15} have 
shown, using weak lensing measurements, that the centre of light is generally significantly offset 
from the true cluster centre while BCGs are, on average, consistent with being at the centre of 
the cluster potential.

We also looked for clusters whose BCG cannot be identified unambiguously because there are other similarly bright member galaxies. Three southern (ACT-CL J0215$-$5212, ACT-CL J0232$-$5257 and ACT-CL J0521$-$5104) and four equatorial (ACT-CL J0239.8$-$0134, ACT-CL J0256.5+0006, ACT-CL J2055.4+0105 and ACT-CL J2302.5+0002) clusters fall under this category \citep[see][for optical images of ACT clusters; \Cref{s:individual} for more detailed comments on some of these clusters]{menanteau10_act,menanteau13}. We estimate the masses of these clusters once more, taking the next most probable BCG candidate (where this is determined by visual inspection) as the cluster centre. In all cases the mass difference is well within the reported uncertainties.

From these two tests we conclude that uncertainties due to the choice of cluster centre are within the quoted errorbars and therefore cluster centring does not introduce any biases or additional uncertainties on our mass estimates.

\section{Notable clusters}
\label{s:individual}

\begin{figure*}
 \centerline{\includegraphics[height=3in]{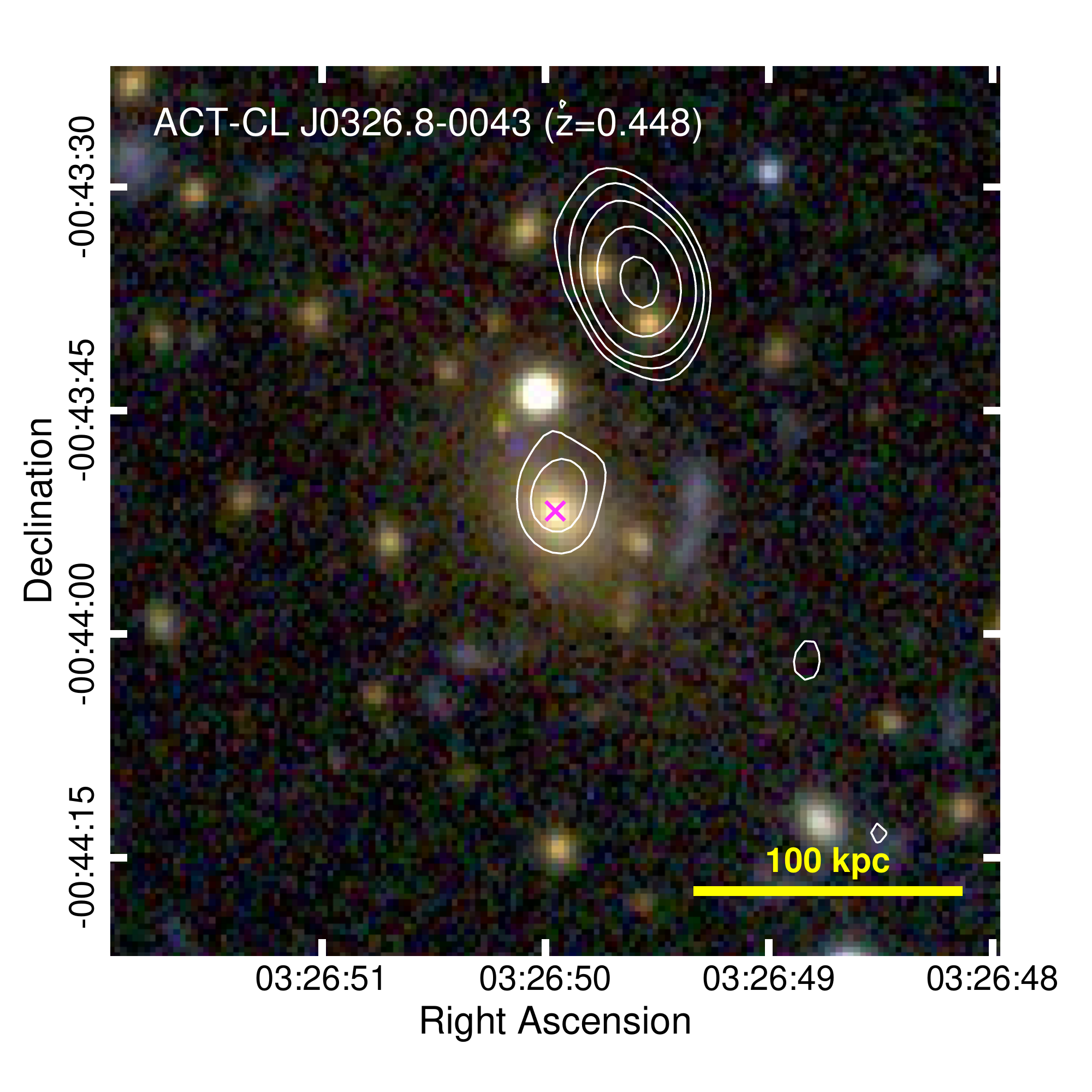}
             \includegraphics[height=3in]{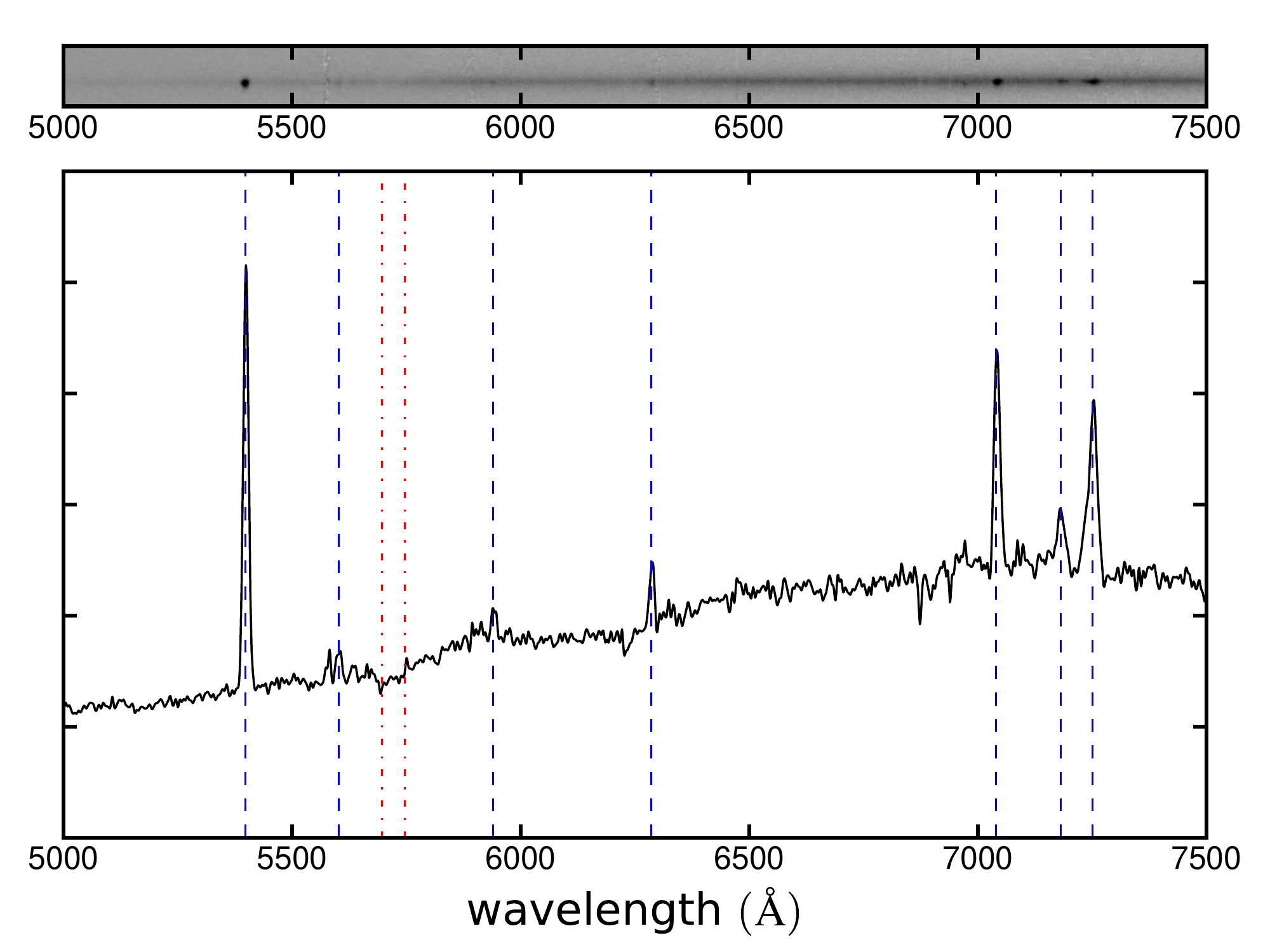}}
\caption{The BCG of ACT-CL~J0326.8$-$0043 at $z=0.45$. \textit{Left:} Optical $gri$ image from 
SDSS with 1.4 GHz contours from the FIRST survey overlaid in white and shown at 
$(3,5,8,15,25)\sigma$ levels, where $\sigma=0.14\,\mathrm{mJy\,beam^{-1}}$. The BCG is marked with 
a magenta cross and the thick yellow line in the bottom-right corner is 100 kpc wide, 
corresponding to $17.\!\arcsec4$ at the cluster redshift. North is up and East is left.
\textit{Right:} Optical one- (bottom) and 
two-dimensional (top) Gemini/GMOS spectra. The former is smoothed with a 3-pixel boxcar. In the 
bottom panel, detected emission lines are marked with dashed blue lines (in order of increasing 
wavelength: [O {\sc ii}], Ne {\sc iii}, H$\delta$, H$\gamma$, H$\beta$ and [O {\sc 
iii}]$\lambda\lambda4959,5007$); the Ca {\sc ii} K,H absorption doublet is marked with dash-dotted 
red lines. The asymmetric broadening of the [O {\sc iii}]$\lambda5007$ line is an artifact 
introduced by interpolating the GMOS chip gaps. The vertical axis is in arbitrary units.}
\label{f:J0326}
\end{figure*}

In this section we describe some notable clusters in the equatorial sample in more detail. Similar 
notes on southern clusters can be found in \cite{sifon13}. We first summarize ACT clusters that 
have been studied in detail elsewhere and then discuss individual clusters.

\subsection{Previously studied ACT clusters}
\label{s:previous}

El Gordo \citep[ACT-CL~J0102$-$4915, $z=0.87$,][]{menanteau12} is probably the most massive cluster known at $z>0.8$ \citep{jee14_gordo}. It is a merging system composed of two roughly equal-mass subclusters colliding approximately perpendicular to the line-of-sight \citep{zitrin13,jee14_gordo}, probably seen about 1 Gyr after core passage \citep{ng15}. It hosts the highest-redshift known radio relics and halo \citep{lindner14}. Its dynamical mass ($M_{200}=(1.13\pm0.29)\times10^{15}\,\Msun$, cf.\ \Cref{t:masses}) is significantly smaller than the total mass estimated from weak lensing \citep[$M_{200}=(2.84\pm0.51)\times10^{15}\,\Msun$,][]{jee14_gordo}, but the former can be expected to be biased when such an extreme system is assumed to be composed of a single component (as is the case here for consistency with the rest of the sample). As a result of the major merger, the total stellar mass in El Gordo is lower than the expectation based on its SZ effect \citep{hilton13}.

ACT-CL~J0022.2$-$0036 ($z=0.81$) is the highest-significance detection in the S82 sample \citep{hasselfield13}. The dynamical mass is consistent with independent mass estimates from weak lensing \citep{miyatake13}, optical richness and high-resolution SZ measurements \citep{reese12}, giving an inverse-variance-weighted average mass of $M_{200}=(7.8\pm0.9)\times10^{14}\Msun$ \citep[see also the discussion in][]{menanteau13}.

ACT-CL~J0256.5+0006 ($z=0.36$) was studied in detail by \cite{knowles16}. It is one of the lowest-mass systems known to host a giant radio halo, which is likely produced by the interaction of two systems with a mass ratio of approximately 2:1 being observed prior to the first core crossing. The merging scenario is supported by the velocity distribution (ACT-CL~J0256.5+0006 has $\mathcal{S}_\Delta<0.01$ at 68 per cent confidence) and X-ray observations; there are two X-ray peaks coincident with two dominant galaxies. The velocity dispersions of the two components suggests that the reported mass, which assumes a single component, may be biased high by roughly 40 per cent, an amount comparable to the quoted uncertainty.

ACT-CL~J0320.4+0032 ($z=0.385$) is one of the few clusters whose BCG is known to host a type II quasar \citep{kirk15}. The low number of observed members precludes a detailed analysis of the cluster structure, but we note that a maximally-predictive histogram \citep{knuth06} of the galaxy velocities shows two peaks and a somewhat asymmetric distribution (see \Cref{f:hist_eq}), which suggests that ACT-CL~J0320.4+0032 is a dynamically young cluster.

\subsection{ACT-CL~J0218.2$-$0041}

ACT-CL~J0218.2$-$0041 ($z=0.673$) is one of the lowest-mass clusters in the sample 
(\Cref{t:masses}). In addition to the cluster itself, we have identified in our spectroscopic data 
an overdensity of eight galaxies at $z=0.82$. Their velocity dispersion, while not necessarily 
representative of this system's velocity dispersion, is $\sigma_{\rm 
gal}=880\,\kms$, which would suggest a mass $M>10^{14}\Msun$. We additionally 
identified a structure of 12 galaxies around $z=0.73$ which, although it only spans 
$6000\,\kms$, has $\sigma_{\rm gal}=2320\,\kms$, suggesting that the 
structure is probably not collapsed.

Two of the three structures have velocity dispersions that suggest cluster-sized systems. The fact 
that we detect these overdensities, instead of non-members being scattered in redshift space, 
suggests that ACT-CL~J0218.2$-$0041 may be associated with a larger cosmic structure along the 
line of sight, with two relatively massive clusters at $z=0.67$ and $z=0.82$ possibly connected by 
a filament (the $z=0.73$ structure). For comparison, we also detected additional galaxy 
overdensities in the lines of sight of ACT-CL~J0235$-$5121 (7 galaxies at $z=0.44$) and 
ACT-CL~J0215.4+0030 (8 galaxies at $z=0.39$), which have $\sigma_{\rm 
gal}=280\,\kms$ and $\sigma_{\rm gal}=170\,\kms$, respectively. This 
shows that low-mass groups may in fact be identified with our observations---that is, the 
structures detected behind ACT-CL~J0218.2$-$0041 are not likely to be low mass groups. This 
large-scale scenario is also appealing given the low mass of ACT-CL~J0218.2$-$0041. It would be 
interesting to explore the impact of this structure on the measured SZ effect, and similarly on 
X-ray emission, but this is deferred to future work.

\subsection{ACT-CL~J0326.8$-$0043}

ACT-CL~J0326.8$-$0043 ($z=0.447$) was first discovered as part of the Massive Cluster Survey 
\citep[MACS~J0326.8$-$0043,][]{ebeling01}. The left panel of \Cref{f:J0326} shows a SDSS $gri$ 
image of the centre of the cluster, with 1.4 GHz contours from the Faint Images of the Radio Sky 
at Twenty centimetres \citep[FIRST,][]{becker95} survey overlaid in white. The BCG (which we refer 
to simply as J0326 in the remainder of this section) shows strong emission lines across the 
optical spectrum (\Cref{f:J0326}, right panel). Because our Gemini/GMOS spectrum is not 
flux-calibrated we use the line measurements from the SDSS MPA/JHU Value-Added Galaxy Catalog 
\citep{brinchmann04} throughout. We cannot distinguish whether the emission in J0326 is 
dominated by star formation or an AGN from the line ratio diagnostic introduced by 
\cite{lamareille10}, appropriate for high-redshift ($z>0.4$) objects for which H$\alpha$ falls 
outside the optical wavelength range (specifically, $\log({\rm [\textrm{O {\sc 
ii}}]}\lambda\lambda3726+3729/{\rm H}\beta)=0.57\pm0.03$ and $\log({\rm [\textrm{O {\sc 
iii}}]}\lambda5007/{\rm H}\beta)=-0.23\pm0.03$).

The left panel of \Cref{f:J0326} shows that there is additionally significant 1.4 GHz emission 
from a point source whose peak is offset $1.\!\arcsec2$ (7 kpc) from the BCG (but note that the 
positional uncertainty in the FIRST source is $\approx1\arcsec$), but again the nature of the 
emission cannot be determined. In the case of star forming galaxies with no AGN contamination, the 
1.4 GHz luminosity can be used as an unobscured tracer of star formation. The 1.4 GHz luminosity 
of the source associated to the BCG is $\log L_{1.4}/(\mathrm{W\,Hz^{-1}})=24.0$, which at face 
value implies a star formation rate (SFR) of several hundred $\Msun\,\mathrm{yr^{-1}}$ 
\citep{hopkins03}. In contrast, the [O {\sc ii}]$\lambda3727$ doublet suggests a SFR of a few tens 
of $\Msun\,\mathrm{yr^{-1}}$. Systems with such marked differences in estimated SFRs are almost 
always AGN hosts (J.\ Brinchmann, private communication). \cite{chang15} fitted spectral energy 
distributions to optical SDSS data plus mid infrared data from the Wide-field Infrared Survey 
Explorer \citep[WISE,][]{wright10} of one million objects. For J0326 they estimated a best-fit star 
formation rate of ${\rm SFR}=15_{-5}^{+10}\,\Msun\,\mathrm{yr^{-1}}$, consistent with the radio 
emission being dominated by nuclear activity. If this is the case then J0326 is a new Type II AGN 
BCG (Type I AGNs are characterized by broad components in the [O {\sc iii}] lines), similar to the 
case of the BCG of ACT-CL~J0320.4+0032 recently reported by \cite{kirk15} and noted in 
\Cref{s:previous}. Therefore J0326 probably adds to the very sparse sample of Type II AGNs in BCGs 
\citep[see references in][]{kirk15}.

\cite{gilmour09} analyzed a 10 ks \textit{Chandra} observation of ACT-CL~J0326.8$-$0043 and found no evidence for an X-ray point source in the BCG location to suggest the presence of an AGN; however, the observations are too shallow to draw any firm conclusions. While available X-ray data are not sufficient to establish the cooling rate in the cluster core, all evidence points to a fairly relaxed cluster. There is no evidence for substructure from the velocity distribution; $\vbcg=205\pm147\,\kms$ suggests the BCG is located at the centre of the potential; and the magnitude gap to the second-brightest member (based on photometric redshifts to avoid a bias due to spectroscopic incompleteness) is relatively large, $\Delta m_{12}=1.62$, which is also an indication of a dynamically old cluster \citep[e.g.,][]{wen13}.

\subsection{ACT-CL~2050.5$-$0055}

\begin{figure}
 \centerline{\includegraphics[width=3.4in]{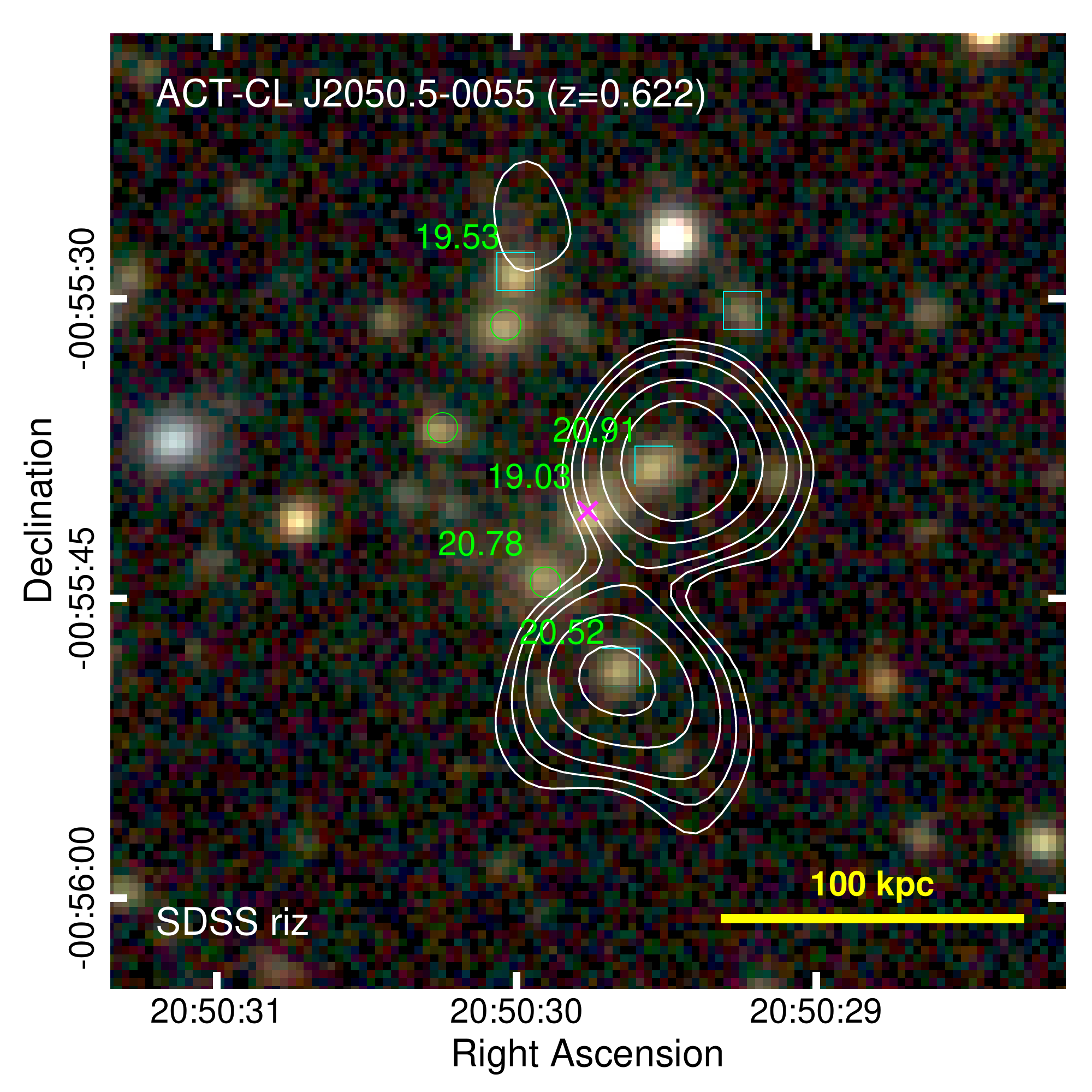}}
\caption{Optical $riz$ image of the central region of ACT-CL~J2050.5$-$0055 ($z=0.62$) from SDSS, 
with 1.4 GHz contours from the FIRST survey overlaid in white at $(3,5,8,15,25)\sigma$ levels, 
where $\sigma=0.15\,\mathrm{mJy\,beam^{-1}}$. The BCG is marked with a magenta cross; cyan squares 
show other spectroscopically confirmed cluster members and green circles show bright photometric 
redshift members. Greeen numbers to the top-left of the five brightest cluster members correspond to dereddened $i$-band magnitudes from SDSS. The thick yellow line in the bottom-right corner is 100 kpc wide, corresponding to $14.\!\arcsec7$ at the cluster redshift. North is up and East is left.}
\label{f:J2050}
\end{figure}

The BCG of ACT-CL~2050.5$-$0055 ($z=0.623$; hereafter simply ``the BCG'') has the highest peculiar velocity of all BCGs in the ACT sample. In fact, the BCG is rejected by our member selection algorithm, with a peculiar velocity of $\vbcg=-(1572\pm143)\,\kms$, different from zero at 11$\sigma$, compared with a cluster velocity dispersion $\sigma_{200}=(511\pm97)\,\kms$, the lowest $\sigma_{200}$ in the entire sample (cf.\ \Cref{t:masses}). The BCG also has a redshift in the SDSS catalogue, $z_{\rm SDSS}=0.6133\pm0.0002$, which would make $\vbcg$ more negative by about $200\,\kms$ (compared to $z_{\rm Gemini}=0.6141\pm0.0003$). For the purpose of this discussion, this difference is not important and we will continue to use $z_{\rm Gemini}$ throughout. Such a high $\vbcg$ probably originated as a result of either merging activity or strong galaxy-galaxy interactions in the centre \citep{martel14}. Regarding the possibility of a cluster-scale merger, the DS test does not reveal any evidence for substructure, although we do not have enough member galaxies to draw firm conclusions. As seen in \Cref{f:substructure}, there are haloes in our Multidark sample that have comparable BCG velocities but they tend to have lower values of $\sdelta$ than ACT-CL~2050.5$-$0055. 

\cite{coziol09} studied a sample of 452 BCGs in low-redshift clusters, and found that BCGs have, 
on average, $\vbcgabs=0.32\sigma$ (where $\sigma$ is the cluster velocity dispersion), with only 
three BCGs having velocities $\vbcgabs>2\sigma$. In comparison, the BCG of ACT-CL~2050.5$-$0055 
has $\vbcg=-(3.1\pm0.7)\,\sigma_{200}$.\footnote{Including the BCG in the member sample by hand 
increases the cluster velocity dispersion to $\sigma_{200}=(607\pm107)\,\kms$, 
yielding $\vbcg=-(2.6\pm0.5)\,\sigma_{200}$.} Similarly, all the BCGs studied by \cite{coziol09} 
have $\vbcgabs<1500\,\kms$. Therefore the BCG of ACT-CL~2050.5$-$0055 is unique in 
this respect; it will be interesting to study the conditions that led to such high $\vbcgabs$. All 
other spectroscopic members in the cluster centre (\Cref{f:J2050}) have peculiar velocities 
between $-550$ and $350\,\kms$, consistent with the low cluster velocity dispersion.

\Cref{f:J2050} shows FIRST contours overlaid on an SDSS $gri$ image of the central region of 
ACT-CL~2050.5$-$0055. There are two point-like sources coinciding with two galaxies within 100 kpc 
from the BCG. The integrated 1.4 GHz luminosities of the northern and southern sources are $\log 
L_{1.4}/(\mathrm{W\,Hz^{-1}})=25.1$ and 25.0, respectively. Such high luminosities suggest that 
indeed these are point sources rather than extended emission originating in the ICM. All 
spectroscopic members shown in \Cref{f:J2050} have spectra typical of passive, elliptical galaxies 
with no signs of activity, with strong Ca {\sc ii} K,H absorption and no optical emission lines.

We also show in \Cref{f:J2050} the deredenned $i$-band magnitudes from the SDSS catalog for the five brightest galaxies (spectroscopic and photometric members). The BCG is brighter than any other galaxy by 0.5 mag. In combination with its central location relative to the galaxy distribution, it is unlikely that the BCG is not the central galaxy of the cluster. In particular, the galaxies that coincide with radio sources are only the third and fifth brightest galaxies, making it unlikely that any of them is the central galaxy. We conclude that misidentification of the central galaxy is unlikely to explain the high $\vbcgabs$ reported for this cluster.

\subsection{ACT-CL~J2055.4+0105}
\label{s:J2055}

As mentioned in \Cref{s:miscentring}, the identification of the BCG is not obvious for this cluster at $z=0.409$. In fact, we identify four galaxies along a straight line extending 1.2 Mpc SE of the BCG (the BCG is the one further NW of the four), the faintest of which is only 0.86 mag fainter than the BCG (all four galaxies are spectroscopically confirmed). Two of the four galaxies (the first and third brightest, and in the NW-SE line joining them, which are separated by 940 kpc) have extended light envelopes characteristic of cD galaxies. Only the fiducial BCG is clearly surrounded by a large overdensity of red galaxies, suggesting it is indeed the central galaxy. The fact that its peculiar velocity is consistent with zero also supports this scenario. The ``alternative BCG'' velocity (of $-700\,\kms$) shown in \Cref{f:hist_eq} corresponds to this secondary cD galaxy; the other two galaxies have peculiar velocities of less than 400 $\kms$.

Overall, this elongated configuration with multiple dominant galaxies strongly suggests that ACT-CL~J2055.4+0105 is undergoing a major merger between at least two massive subclusters, and possibly more. The DS test does not reveal any evidence for substructure ($\sdelta=0.81\pm0.10$), but we note that this is also the case for El Gordo \citep{sifon13}. This is because the DS test (nor, indeed, dynamical information in general) is not sensitive to mergers happening along the plane of the sky. As with other clusters described in this section, more data would be required for a detailed assessment of this system.

\subsection{ACT-CL~J2302.5+0002}
\label{s:J2302}

\begin{figure}
 \centerline{\includegraphics[width=3.2in]{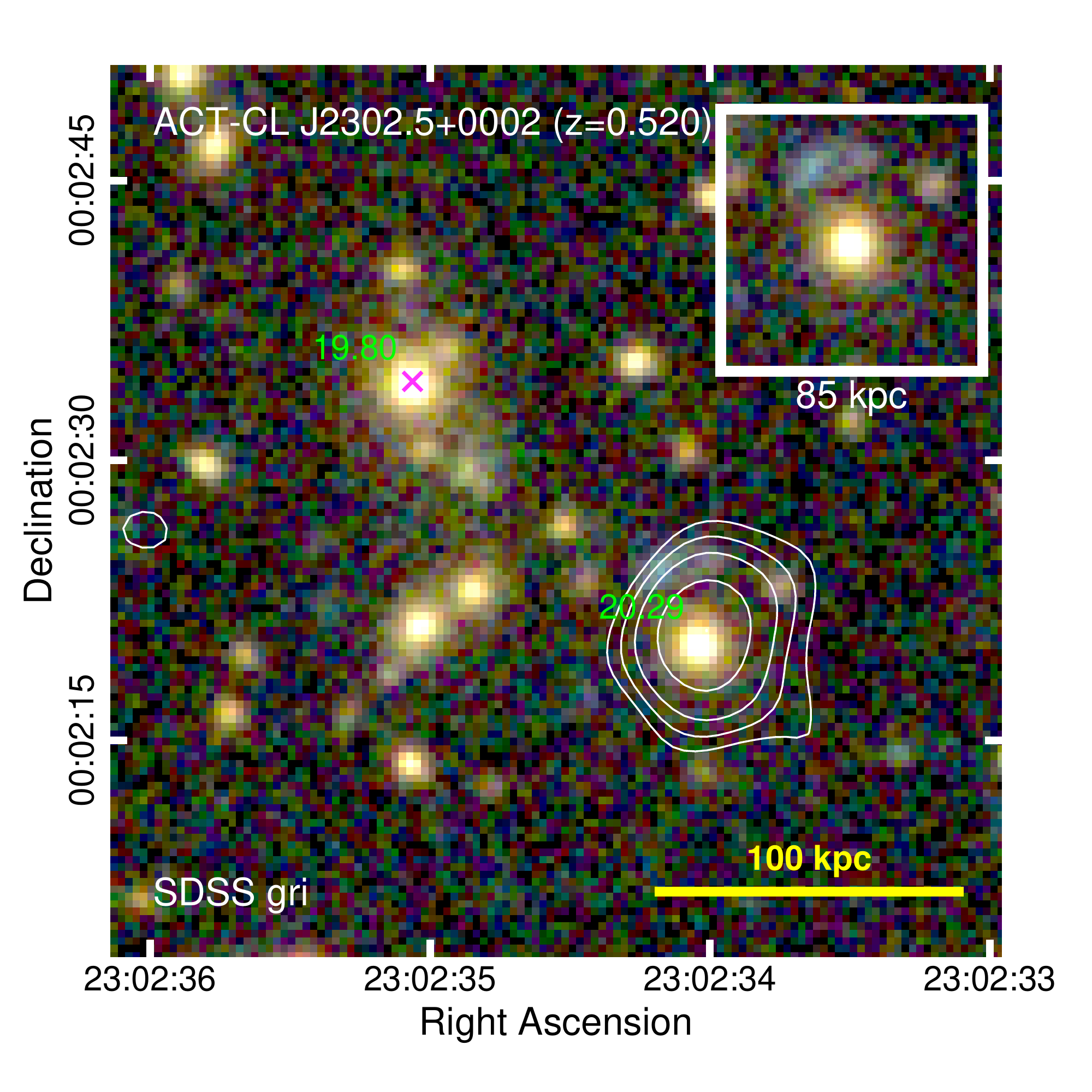}}
\caption{Optical SDSS $gri$ image of the inner 300 kpc of ACT-CL J2302.5+0002. Symbols, labels and contours are as in \Cref{f:J2050}. The inset in the top-right corner is 85 kpc on a side and shows the region around the second-brightest galaxy without the radio contours, which show what appears to be a strongly lensed background image. The thick yellow line in the bottom-right corner is 100 kpc wide, corresponding to $16.\!\arcsec0$ at the cluster redshift. North is up and East is left.}
\label{f:J2302}
\end{figure}

ACT-CL~J2302.5+0002 at $z=0.520$ is also one of the lowest-mass clusters in the sample. We show the central 300 kpc of this cluster in \Cref{f:J2302}, which shows strong evidence that the brightest galaxy is not the \emph{central} cluster galaxy: the second-brightest galaxy, 130 kpc away, shows strong 1.4 GHz emission and what appears to be a strongly lensed background galaxy around it. The coordinates given by \cite{menanteau13} indeed correspond to the brightest galaxy, but for our purposes we take the second-brightest galaxy as the cluster centre. ACT-CL J2302.5+0005 is therefore the only cluster for which we do not take the brightest galaxy as the cluster centre in \Cref{s:masses}. As mentioned in \Cref{s:miscentring}, adopting either galaxy as the centre gives consistent mass estimates; using the brightest galaxy instead of the central one we obtain $M_{200}=(1.9\pm0.7)\times10^{14}\Msun$, compared to the fiducial value of $M_{200}=(2.4\pm0.7)\times10^{14}\Msun$. We do not detect any optical emission lines in either galaxy.

Although such clear examples of brightest galaxies not being the central galaxy are rare, the value of the separation between the two galaxis is not uncommon \citep{skibba11,martel14}. \cite{martel14} have argued that the positional offset of the BCG is not as robust an indication of major mergers as the velocity offset, $\vbcg/\sigma$, where $\sigma$ is the cluster velocity dispersion. As shown in \Cref{f:hist_eq}, both galaxies have large peculiar velocities, with $v=-470\,\kms$ ($v/\sigma_{200}=0.70$) and $v=-600\,\kms$ ($v/\sigma_{200}=0.89$) for the central and brightest galaxies, respectively. Similar to (but less extreme than) the BCG in ACT-CL J2050.5$-$0055, these peculiar velocities---especially given the small cluster velocity dispersion---strongly suggest that ACT-CL~J2302.5+0002 is undergoing a major merger \citep{martel14}, but the available data do not allow us to perform a more detailed analysis.

\section{Conclusions}\label{s:conclusions}

We have carried out a spectroscopic follow-up effort of ACT SZ-selected clusters in the equatorial 
survey in the redshift range $0.3<z<0.9$. Combined with our previous follow-up program of the 
southern clusters and archival data, we present velocity dispersions and dynamical masses for 44 
clusters at $0.24<z<1.06$ with a median of 55 spectroscopic members per cluster.

We calibrate our velocity dispersion measurements using the Multidark simulation, taking into 
account the spectroscopic coverage of clusters which is qualitatively different for southern and 
equatorial clusters, owing to the different optical imaging available (namely, targeted 
$5\arcmin\times5\arcmin$ observations in the south and full SDSS coverage in the equator, 
\Cref{f:observations}). We find that velocity dispersions are unbiased so long as the measurement 
includes galaxies out to $r_{200}$ but the azimuthal distribution of spectroscopic targets is not 
important for our purposes (see \Cref{s:sims} and \Cref{f:veldisp_sims}). We use the average radial 
velocity dispersion profile of subhaloes in Multidark to correct measurements for clusters whose 
coverage does not reach $r_{200}$ and include the uncertainties from this correction 
(\Cref{t:MDcorr}) in the reported velocity dispersions.

We use a scaling relation between galaxy velocity dispersion and cluster mass derived from zoomed 
cosmological hydrodynamical simulations to infer dynamical masses consistently for the full sample 
of clusters with spectroscopic observations. We make a detailed assessment of the different 
contributions to the reported mass uncertainties, which are dominated by a $\approx$30 per cent 
scatter in the scaling relation induced by interlopers, triaxiality and the intrinsic scatter of 
the relation. Because this is a constant value, we do not include this contribution in the 
reported cluster mass uncertainties but recommend that it be included in cosmological analyses 
derived from these data. Statistical uncertainties from our average 55 members per cluster are 
comparable to said scatter, while uncertainties from member selection and the spectroscopic 
aperture selection are subdominant.

The updated dynamical mass estimates of the southern clusters are, on average, 71 per cent of the masses presented in \cite{sifon13}. This overall difference results from (i) accounting for the observing strategies used to get the galaxy redshifts when calculating cluster velocity dispersions (\Cref{s:sims}), and (ii) using a $\sigma-M_{200}$ scaling relation that includes the effects of baryonic physics and dynamical friction (\Cref{s:scaling}). We find that masses derived from the SZ effect assuming a scaling relation based on the pressure profile of \cite{arnaud10} are consistent with the dynamical masses and report a mass bias which results from the combination of the dynamical mass bias and the SZ bias, $(1-b_{\rm SZ})/\beta_{\rm dyn} = \avg{M}_{500}^{\rm SZ} / \avg{M}_{500}^{\rm dyn}=\massbias$, with an additional 0.14 systematic uncertainty due to the unknown galaxy velocity bias (see \Cref{f:msz} and \Cref{s:msz}), consistent with previous estimates from the literature if one accounts for the different mass regimes. \cite{hasselfield13} used the dynamical masses of \cite{sifon13} as prior information on the cosmological analysis derived from the SZ cluster counts and found that dynamical masses suggested a higher $\sigma_8$ than other cosmological probes. The new, lower dynamical masses will bring the estimate of $\sigma_8$ down. A cosmological analysis incorporating these new dynamical mass estimates will be presented in a future paper.

We also highlight five newly-characterized clusters. ACT-CL J0218.2$-$0041 ($z=0.67$) appears to be part of a structure where two cluster-sized systems are connected by a filament along the line of sight. The BCG of ACT-CL J0326.8$-$0043 ($z=0.45$) is likely a rare Type II AGN host which also seems to be associated with strong radio emission. The BCG of ACT-CL 2050.5$-$0055 ($z=0.62$) has a peculiar velocity of $\vbcg=3\sigma_{200}$ and is surrounded by double-peaked (probably point-source) radio emission. ACT-CL~J2055.4+0105 ($z=0.41$) has four bright, locally-dominant galaxies separated by 1.2 Mpc along a straight line. Finally, ACT-CL~J2302.5+0002 ($z=0.52$) is a clear example of the brightest galaxy not being the central cluster galaxy, confirmed by the presence of a strong lensing arc around the second-brightest galaxy. Further follow-up studies will reveal more details about these intriguing systems.

The uncertainty on the average dynamical mass is dominated by the scatter in the $\sigma-M$ relation (see \Cref{t:uncertainties}), which cannot be significantly reduced by observing more galaxies per cluster. \cite{ntampaka15} recently developed a machine learning approach to measure dynamical masses which incorporates information about the distribution of galaxies and their velocities to predict cluster masses, which has been successfully applied to mock observations that include the effects of impurity and incompleteness, reducing the errors by up to 60 per cent \citep{ntampaka16}. Further tests on more realistic galaxy catalogues will assess its effectiveness in measuring galaxy cluster masses in real observations.

\section*{Acknowledgments}

We thank Jarle Brinchmann for help interpreting the spectrum of the BCG of ACT-CL~J0326.8$-$0043, Gary Mamon for useful discussions on cluster velocity dispersion profiles and Stefano Andreon and Bruce Partridge for useful comments during the preparation of this manuscript. We are grateful to the referee, August Evrard, for a thorough review of our work, which helped improve the clarity and robustness of the discussions presented in this paper.

C.S.\ acknowledges support from the European Research Council under FP7 grant number 279396 awarded 
to H.\ Hoekstra.
N.B.\ and R.H.\ acknowledge support from the Spitzer Fellowship.
L.F.B.'s research is supported by FONDECYT under project 1120676.
A.K.\ acknowledges support from NSF grant 1312380.

This work was supported by the U.S. National Science Foundation through awards AST-0408698 and 
AST-0965625 for the ACT project, as well as awards  PHY-0855887 and PHY-1214379. Funding was also 
provided by Princeton University, the University of Pennsylvania, and a Canada Foundation for 
Innovation (CFI) award to UBC. ACT operates in the Parque Astron\'omico Atacama in northern Chile 
under the auspices of the Comisi\'on Nacional de Investigaci\'on Cient\'ifica y Tecnol\'ogica de 
Chile (CONICYT). Computations were performed on the GPC supercomputer at the SciNet HPC Consortium. 
SciNet is funded by the CFI under the auspices of Compute Canada, the Government of Ontario, the 
Ontario Research Fund -- Research Excellence; and the University of Toronto.

Based in part on observations  obtained at the Gemini Observatory, which is operated by the 
Association of Universities for Research in Astronomy, Inc., under a cooperative agreement with 
the NSF on behalf of the Gemini partnership: the National Science Foundation (United States), the 
Science and Technology Facilities Council (United Kingdom), the National Research Council 
(Canada), CONICYT (Chile), the Australian Research Council (Australia), Minist\'erio da Ci\^encia 
e Tecnologia (Brazil) and Ministerio de Ciencia, Tecnolog\'ia e Innovaci\'on Productiva 
(Argentina).

Funding for SDSS-III has been provided by the Alfred P. Sloan Foundation, the Participating 
Institutions, the National Science Foundation, and the U.S. Department of Energy Office of 
Science. The SDSS-III web site is \url{http://www.sdss9.org/}.

The MultiDark Database used in this paper and the web application providing online access to it 
were constructed as part of the activities of the German Astrophysical Virtual Observatory as 
result of a collaboration between the Leibniz-Institute for Astrophysics Potsdam (AIP) and the 
Spanish MultiDark Consolider Project CSD2009-00064. The Bolshoi and MultiDark simulations were run 
on the NASA's Pleiades supercomputer at the NASA Ames Research Center. The MultiDark-Planck (MDPL) 
and the BigMD simulation suite have been performed in the Supermuc supercomputer at LRZ using time 
granted by PRACE.

IRAF is distributed by the National Optical Astronomy Observatory, which is operated by the 
Association of Universities for Research in Astronomy (AURA) under a cooperative agreement with 
the National Science Foundation.

This work has made use of \textsc{IPython} \citep{perez07} and of the python packages 
\textsc{numpy} and \textsc{scipy}. Plots have been made using \textsc{matplotlib} \citep{hunter07}.

\bibliographystyle{mnras}
\bibliography{bibliography}

\begin{appendix}

\section{Eddington bias and selection effects}
\label{ap:bias}

\begin{figure}
 \centerline{\includegraphics[width=3.2in]{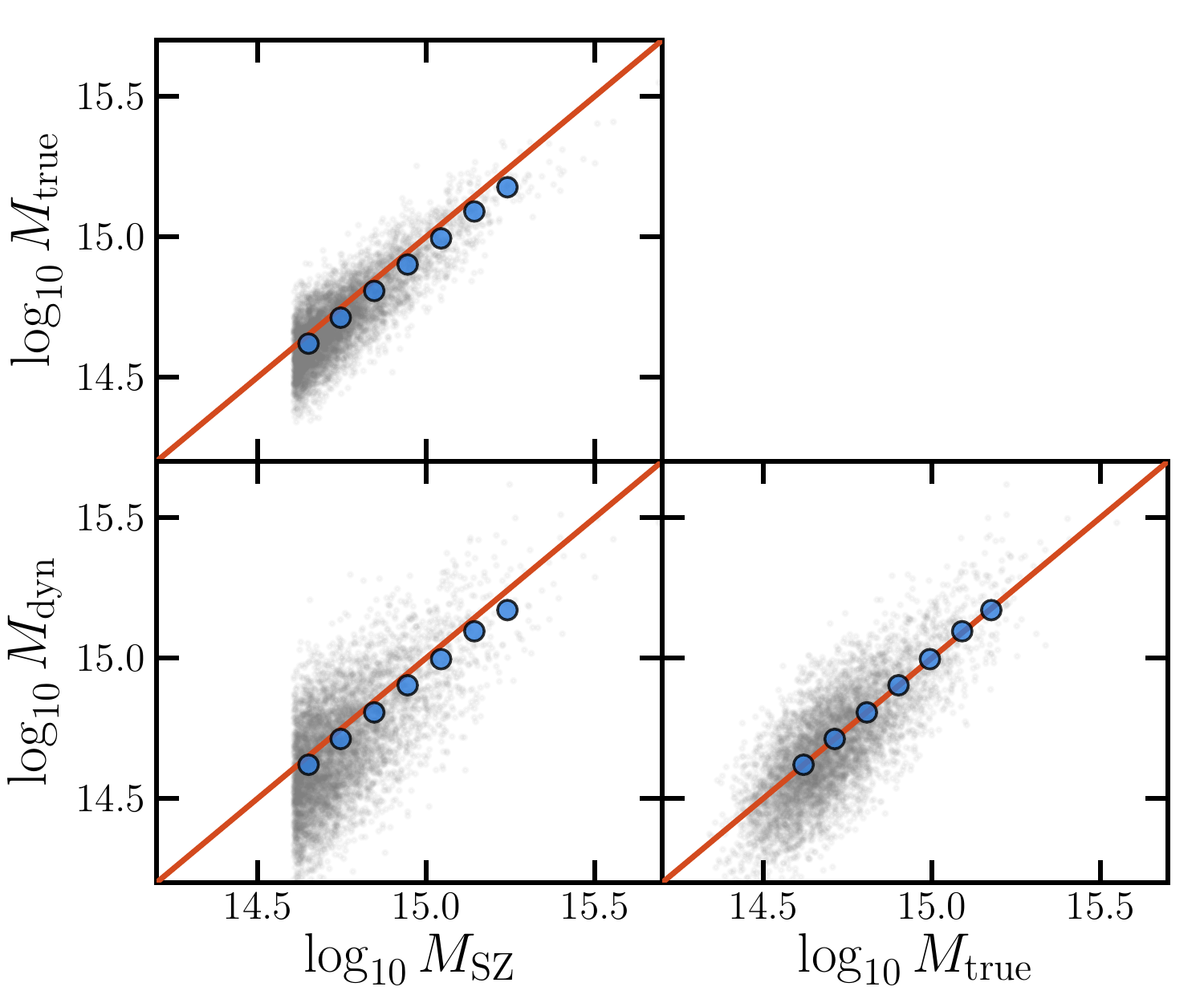}}
\caption{Comparison of SZ, dynamical, and true cluster masses for a mock catalog that incorporates 
a simple model for noise, scatter and selection effects.  The gray points are individual mock 
clusters. The blue points show the mean masses for several bins in $M_{\rm SZ}$. Because the 
binning is based on the measured $M_{\rm SZ}$ value, the binned $M_{\rm SZ}$ are consistently 
higher than the binned $M_{\rm true}$ (upper left panel).  In contrast, the $M_{\rm dyn}$ are not 
affected by Eddington bias (lower right panel) and lie on the $y=x$ (red) line.}
\label{f:appendix_a1}
\end{figure}

\begin{figure}
 \centerline{\includegraphics[width=3.2in]{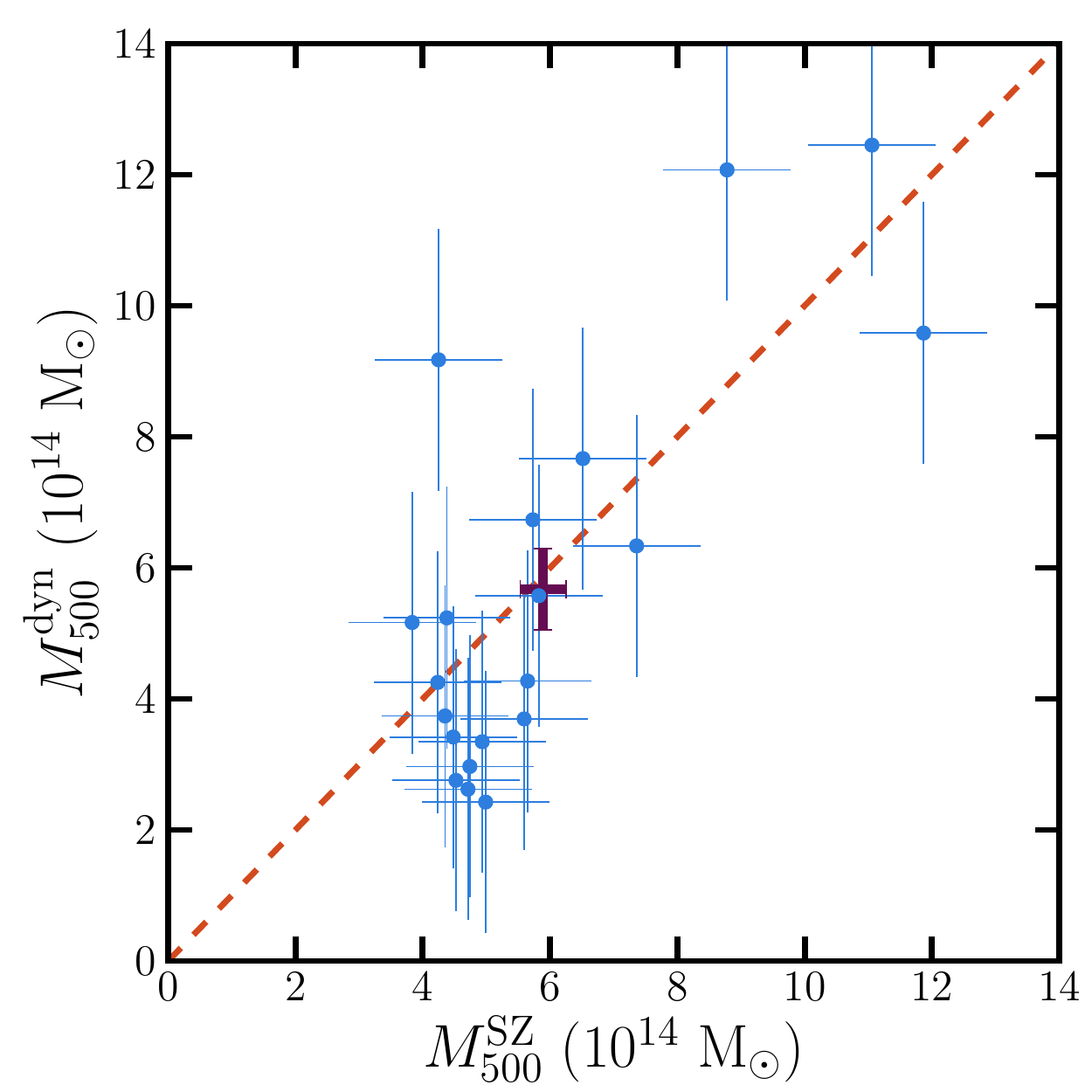}}
\caption{Masses of 20 clusters drawn from a large mock catalog that simulates noise, scatter, and 
selection effects. Purple cross shows the mean characteristic SZ and dynamical mass, computed under 
uniform weights. As in \Cref{f:msz}, errorbars in the individual cluster masses include only the 
noise contribution and not intrinsic scatter.}
\label{f:appendix_a2}
\end{figure}

In this appendix we discuss the SZ and dynamical mass measurements in the context of understanding potential biases, such as Eddington bias \citep{eddington13}, that might affect the comparison of the two mass proxies. We then demonstrate with a simple simulation that the method used in \Cref{s:mdyn} to determine the ratio $(1-b_{\rm SZ})/\beta_{\rm dyn}$ is unbiased.

The SZ masses used in this work have been corrected for Eddington bias by \cite{hasselfield13}. The calculation accounts for the fact that
the underlying mass function is falling steeply as mass increases, so a direct inversion of $y_{\rm SZ}(M)$ gives a biased estimate of $M(y_{\rm SZ})$ \citep{evrard14}. The computation of the correction requires certain assumptions, such as the form of the cluster mass function and the survey selection function, which is constructed assuming a particular model for the SZ signal and an assumption about the degree of intrinsic scatter. Under these assumptions the correction is then computed in a Bayesian framework, with the posterior probability of a cluster's mass given by $P(M \vert y) \propto P(y \vert M) n(M)$, where the likelihood $P(y \vert M)$ accounts for intrinsic scatter and measurement noise and the prior $n(M)$ is the cluster mass function
\citep[see section 3.2 of][]{hasselfield13}.

Aside from the modelling assumptions, it is important to note that the correction is no longer valid if we change the underlying distribution of cluster masses upon which the ACT selection is effectively acting. For example, if we were to remove objects from the sample based on some auxiliary information (such as X-ray flux, or membership in an optical cluster survey), then we would risk complicating the underlying mass function and invalidating the Eddington bias correction applied in \cite{hasselfield13}.

However, there are many ways to sub-sample the ACT sample without changing the validity of the Eddington bias correction. Restricting the survey to a smaller region of the sky, or to a particular redshift range\footnote{The redshift cut can in principle affect the bias correction due to uncertainty in the cluster redshifts; in practice the redshift uncertainties are small enough that this is not significant.} does not affect the Eddington bias correction. Although less obvious, it is also true that raising the S/N threshold of the catalog does not change the correction (though of course this will change the membership of the sample). While the S/N threshold affects the survey selection function (which we take to mean the probability that a cluster of some mass and redshift would be included in the sample), it does not affect the underlying distribution of masses which we should consider when working with a particular cluster that has been detected. In a Bayesian framework, the posterior distribution for any one cluster's mass is not dependent on the overall survey selection function. The cosmological sample considered in this work satisfies the requirements above, and so we take the masses of \cite{hasselfield13}, which have been corrected for Eddington bias, to be unbiased on average.

The dynamical masses presented in this work were obtained for all clusters passing certain redshift and S/N cuts \citep[i.e., those in the cosmological sample of][]{hasselfield13}, and the present measurements were not used to refine the sample further.  Though noise and scatter will certainly affect the velocity dispersion measurements, we expect positive and negative noise (or scatter) excursions to be equally likely. The dynamical mass measurements thus constitute a complete set of ``follow-up'' observations for the sample, and are not affected by Eddington bias.

While the descriptions above can be justified formally \citep[see][]{evrard14}, we illustrate their validity using mock catalogs of SZ and dynamical mass measurements.  We draw a large number of masses, $M_{\rm true}$, from a realistic cluster mass function, considering a single redshift.  We create SZ and dynamical mass proxy measurements by adding intrinsic scatter (at the 20 and 30 per cent levels, respectively) and measurement noise (fixed at $10^{14}\,M_\odot$ and $2\times 10^{14}\,M_\odot$, respectively) to the true masses.  Then we simulate the effect of SZ selection by keeping only mock clusters with $M_{\rm SZ} > 4\times 10^{14} M_\odot$.

In \Cref{f:appendix_a1} we show these mock clusters, and demonstrate that if we bin the objects according to their $M_{\rm SZ}$ measurement, we see Eddington bias effects in $M_{\rm SZ}$ relative to the true cluster mass, but $M_{\rm dyn}$ is not biased.  In the real observations, the Eddington bias in $M_{\rm SZ}$ is corrected by modeling the mass function and SZ scaling relation.  For the mock study we simply fit a constant bias factor to the binned $M_{\rm SZ}$ and $M_{true}$ points and use it to correct the individual mock $M_{\rm SZ}$ values.

We then take a random subsample consisting of 20 mock clusters, to roughly match the size of the real sample considered in this work.  We compute the characteristic SZ and dynamical mass of these clusters, under uniform weights, as was done in \Cref{s:msz}.  The resulting ratio is consistent, as expected, with unity.  This subsample and the characteristic masses are plotted in \Cref{f:appendix_a2}.

We also repeat the entire procedure for alternative weighting schemes. We find that weighting 
schemes that incorporate the measured dynamical masses are significantly biased. For example, if 
we take the weights to be the inverse square of the combined measurement error and intrinsic 
scatter, the characteristic dynamical mass is 10 per cent lower, on average, than the 
characteristic SZ mass. Such biases are also seen in weights that include the fractional 
measurement error (which requires the input of the measured mass), or weights that incorporate the 
intrinsic scatter contribution (which scales in proportion to the measured mass). The real data 
contain additional correlations between dynamical mass and measurement error beyond those modeled 
in this simple simulation, because more massive clusters also tend to have more galaxies that can 
be used to measure the dispersion.  For these reasons it is clear that one should not incorporate 
the dynamical mass measurements and errors into the weights unless the impact can be fully 
modelled.

In constrast, for this simulation we find that weighting by the inverse square of the SZ mass 
error (including the contribution from intrinsic scatter) does not bias the mass comparison. This 
is because we have already corrected the SZ masses to make them unbiased, even under weights (or 
selection choices) that depend on the measured SZ mass.

\section{Velocity histograms and redshift catalogue}

\begin{figure*}
\centerline{\includegraphics[width=7in]{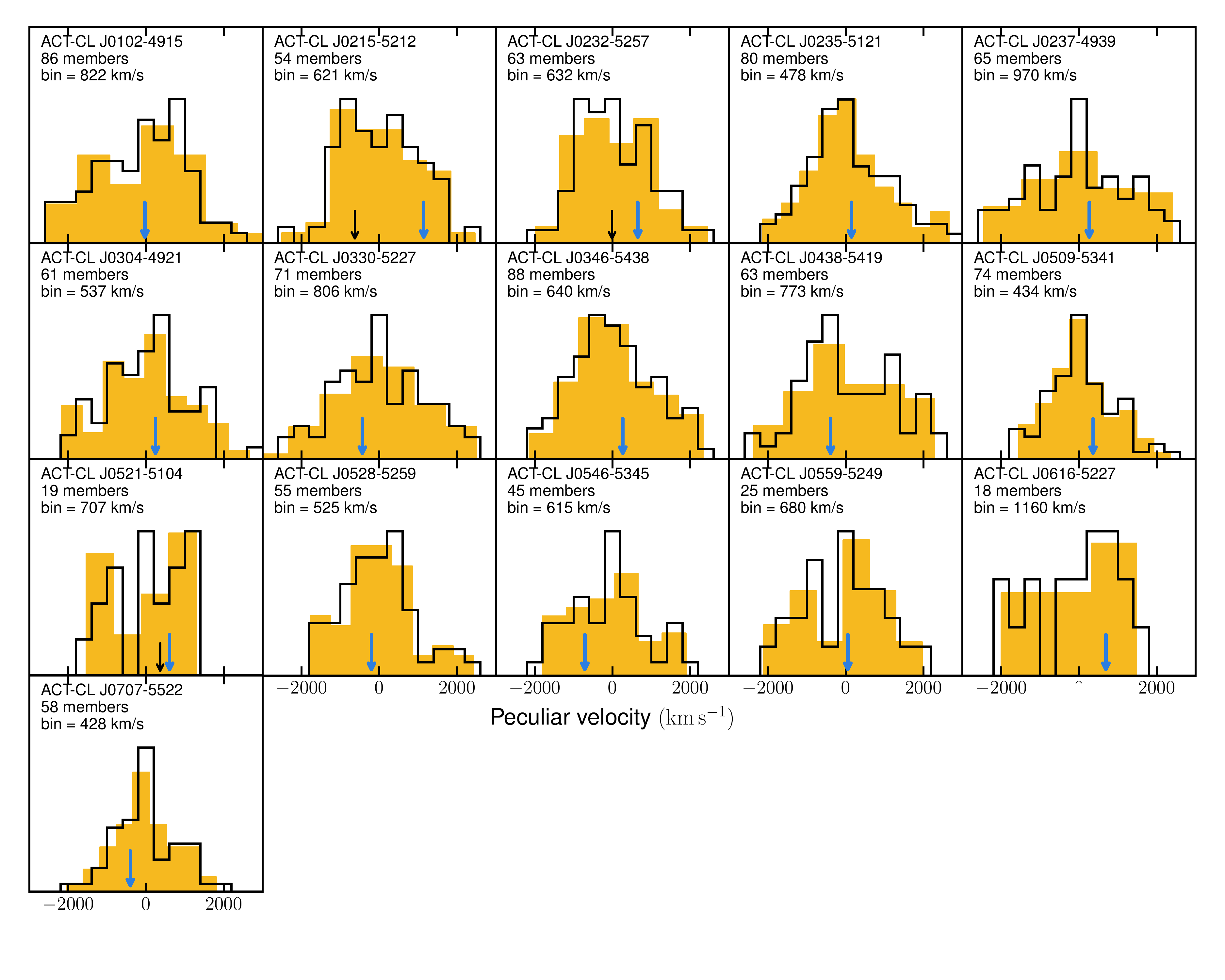}}
\caption{Velocity histograms of all clusters in the southern sample. Black empty 
histograms have a bin size of $400\,\kms$ and filled yellow histograms have a bin 
size indicated in the legend, which is such that the predictive power of the histogram (i.e., the 
likelihood that the next datum will fall in a given bin) is maximized using a bayesian approach. 
We show normalized counts and list the number of members in each cluster in the legends (see also 
\Cref{t:masses}). Blue arrows mark the BCG velocities and, where applicable, smaller black arrows mark peculiar velocities of alternative choices for the BCG (see \Cref{s:miscentring,s:individual}). BCG velocities have a typical uncertainty of $\approx100\,\kms$.}
\label{f:hist_south}
\end{figure*}

\begin{figure*}
\centerline{\includegraphics[width=7in]{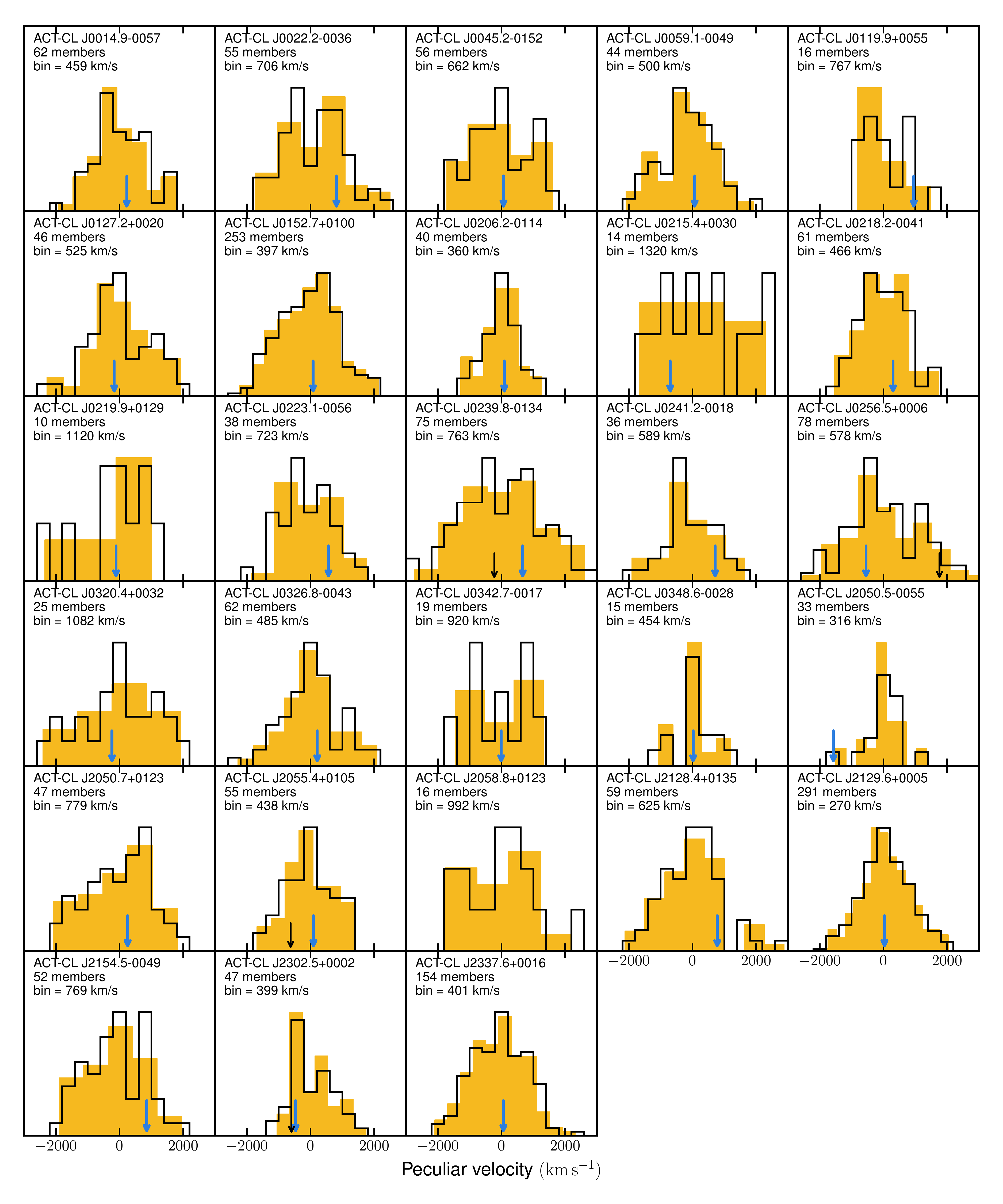}}
\caption{Rest-frame velocity histograms of all clusters in the equatorial sample. Styles are the same as \Cref{f:hist_eq}. We do not know the redshift of the BCG of ACT-CL~J2058.8+0123.}
\label{f:hist_eq}
\end{figure*}

We show in \Cref{f:hist_eq,f:hist_south} the velocity histograms for all clusters (only member 
galaxies are shown). We show histograms with bins of $400\,\kms$ and with a 
constant bin size that maximizes the predictive power of the histogram in a Bayesian sense 
\citep{knuth06}\footnote{We use the version implemented in the \texttt{plotting} library of 
\texttt{astroML} \citep{vanderplas12}.}.

\begin{table*}
\centering
\caption{BCGs of the 44 ACT SZ-selected clusters studied in this work. The full spectroscopic catalog will be available in the online version of the journal. A portion is shown here for guidance regarding its form and content. Columns are: (1): catalogue designation (BCGs are marked with asterisks); (2), (3): J2000 coordinates; (4), redshift and nominal uncertainty; (5): membership flag (1: member, 0: non-member); (6): spectroscopic source. References are: (1) Gemini/GMOS (this work), (2) Gemini/GMOS \citep{sifon13}, (3) VLT/FORS2 \citep{sifon13}, (4) SALT/RSS \citep{kirk15}, (5) SDSS DR12 \citep{alam15}, (6) HeCS \citep{rines13}, (7) \citet{soucail88}, (8) \citet{dressler99}.}
\label{t:redshifts}
\begin{tabular}{l c c c c c}
\hline\hline
 Name & RA & Decl. & $z$ & Member? & Source \\[0.2ex]
\hline
J001454.10$-$005708.4$^*$ & 00:14:54.10 & $-$00:57:08.4 & $0.53435\pm0.00035$ & 1 & 1 \\
J002213.03$-$003633.8$^*$ & 00:22:13.03 & $-$00:36:33.8 & $0.80960\pm0.00031$ & 1 & 1 \\
J004512.49$-$015231.6$^*$ & 00:45:12.49 & $-$01:52:31.6 & $0.54862\pm0.00023$ & 1 & 1 \\
J005806.10$+$003050.0$^*$ & 00:58:06.10 & $+$00:30:50.0 & $0.79223\pm0.00029$ & 1 & 1 \\
J005908.50$-$005005.7$^*$ & 00:59:08.50 & $-$00:50:05.7 & $0.78740\pm0.00018$ & 1 & 1 \\
J011958.14$+$005533.6$^*$ & 01:19:58.14 & $+$00:55:33.6 & $0.73690\pm0.00015$ & 1 & 1 \\
J012716.64$+$002040.9$^*$ & 01:27:16.64 & $+$00:20:40.9 & $0.37933\pm0.00023$ & 1 & 1 \\
J015241.95$+$010025.5$^*$ & 01:52:41.95 & $+$01:00:25.5 & $0.22953\pm0.00012$ & 1 & 6 \\
J015624.29$-$012317.3$^*$ & 01:56:24.29 & $-$01:23:17.3 & $0.45260\pm0.00039$ & 1 & 4 \\
J020613.14$-$011500.0$^*$ & 02:06:13.14 & $-$01:15:00.0 & $0.67629\pm0.00014$ & 1 & 1 \\
J021527.94$+$003050.0$^*$ & 02:15:27.94 & $+$00:30:50.0 & $0.85786\pm0.00031$ & 1 & 1 \\
J021816.88$-$004141.8$^*$ & 02:18:16.88 & $-$00:41:41.8 & $0.67448\pm0.00020$ & 1 & 1 \\
J021952.15$+$012952.1$^*$ & 02:19:52.15 & $+$01:29:52.1 & $0.36460\pm0.00039$ & 1 & 4 \\
J022310.05$-$005709.0$^*$ & 02:23:10.05 & $-$00:57:09.0 & $0.66633\pm0.00015$ & 1 & 1 \\
J023952.74$-$013418.9$^*$ & 02:39:52.74 & $-$01:34:18.9 & $0.37815\pm0.00000$ & 1 & 7 \\
J024115.44$-$001841.0$^*$ & 02:41:15.44 & $-$00:18:41.0 & $0.69121\pm0.00029$ & 1 & 1 \\
J025633.76$+$000628.8$^*$ & 02:56:33.76 & $+$00:06:28.8 & $0.36001\pm0.00013$ & 1 & 1 \\
J032029.78$+$003153.6$^*$ & 03:20:29.78 & $+$00:31:53.6 & $0.38360\pm0.00039$ & 1 & 4 \\
J032649.95$-$004351.7$^*$ & 03:26:49.95 & $-$00:43:51.7 & $0.44812\pm0.00004$ & 1 & 1 \\
J034242.65$-$001708.3$^*$ & 03:42:42.65 & $-$00:17:08.3 & $0.30716\pm0.00005$ & 1 & 5 \\
J034839.54$-$002816.8$^*$ & 03:48:39.54 & $-$00:28:16.8 & $0.34500\pm0.00039$ & 1 & 4 \\
J205029.76$-$005540.6$^*$ & 20:50:29.76 & $-$00:55:40.6 & $0.61409\pm0.00032$ & 0 & 1 \\
J205043.13$+$012329.2$^*$ & 20:50:43.13 & $+$01:23:29.2 & $0.33503\pm0.00036$ & 1 & 1 \\
J205523.25$+$010607.5$^*$ & 20:55:23.25 & $+$01:06:07.5 & $0.40933\pm0.00023$ & 1 & 1 \\
J212823.42$+$013536.4$^*$ & 21:28:23.42 & $+$01:35:36.4 & $0.38920\pm0.00012$ & 1 & 1 \\
J212939.96$+$000521.1$^*$ & 21:29:39.96 & $+$00:05:21.1 & $0.23393\pm0.00013$ & 1 & 6 \\
J215432.35$-$004900.4$^*$ & 21:54:32.35 & $-$00:49:00.4 & $0.49454\pm0.00027$ & 1 & 1 \\
J230235.05$+$000234.2$^*$ & 23:02:35.05 & $+$00:02:34.2 & $0.51666\pm0.00018$ & 1 & 1 \\
J233739.72$+$001616.9$^*$ & 23:37:39.72 & $+$00:16:16.9 & $0.27715\pm0.00018$ & 1 & 6 \\
J010257.74$-$491619.2$^*$ & 01:02:57.74 & $-$49:16:19.2 & $0.86991\pm0.00030$ & 1 & 3 \\
J021512.26$-$521225.2$^*$ & 02:15:12.26 & $-$52:12:25.2 & $0.48595\pm0.00016$ & 1 & 2 \\
J023249.47$-$525711.1$^*$ & 02:32:49.47 & $-$52:57:11.1 & $0.55948\pm0.00034$ & 1 & 2 \\
J023545.28$-$512105.0$^*$ & 02:35:45.28 & $-$51:21:05.0 & $0.27813\pm0.00015$ & 1 & 2 \\
J023701.71$-$493809.9$^*$ & 02:37:01.71 & $-$49:38:09.9 & $0.33548\pm0.00016$ & 1 & 2 \\
J030416.03$-$492126.2$^*$ & 03:04:16.03 & $-$49:21:26.2 & $0.39283\pm0.00020$ & 1 & 2 \\
J033056.83$-$522813.8$^*$ & 03:30:56.83 & $-$52:28:13.8 & $0.43961\pm0.00019$ & 1 & 2 \\
J034655.49$-$543854.9$^*$ & 03:46:55.49 & $-$54:38:54.9 & $0.53105\pm0.00013$ & 1 & 2 \\
J043817.70$-$541920.6$^*$ & 04:38:17.70 & $-$54:19:20.6 & $0.41935\pm0.00012$ & 1 & 2 \\
J050921.38$-$534212.2$^*$ & 05:09:21.38 & $-$53:42:12.2 & $0.46196\pm0.00022$ & 1 & 3 \\
J052114.54$-$510418.4$^*$ & 05:21:14.54 & $-$51:04:18.4 & $0.67719\pm0.00041$ & 1 & 2 \\
J052805.30$-$525952.8$^*$ & 05:28:05.30 & $-$52:59:52.8 & $0.76632\pm0.00037$ & 1 & 3 \\
J054637.67$-$534531.3$^*$ & 05:46:37.67 & $-$53:45:31.3 & $1.06190\pm0.00016$ & 1 & 3 \\
J055943.23$-$524927.1$^*$ & 05:59:43.23 & $-$52:49:27.1 & $0.60969\pm0.00027$ & 1 & 3 \\
J061634.05$-$522710.0$^*$ & 06:16:34.05 & $-$52:27:10.0 & $0.68759\pm0.00011$ & 1 & 2 \\
J070704.67$-$552308.5$^*$ & 07:07:04.67 & $-$55:23:08.5 & $0.29405\pm0.00019$ & 1 & 2 \\
\hline
\end{tabular}
\end{table*}

\Cref{t:redshifts} presents the redshift measurements for all galaxies that enter the analysis. We 
include both members and non-members to allow for reproducibility and independent analyses to be 
carried out with these data. All redshifts are in the heliocentric frame. Note that this is not 
the case in the redshift catalogue published by \cite{sifon13}. We therefore include redshifts for 
galaxies in the southern survey here for consistency.

\end{appendix}

\bsp

\end{document}